\documentclass{article}

\usepackage{arxiv}

\usepackage{minted}

\usepackage{amsmath,amsfonts,graphicx,amssymb,bm,amsthm}
\usepackage{algorithm,algorithmicx}
\usepackage[noend]{algpseudocode}
\usepackage{fancyhdr}
\usepackage{mathtools}

\usepackage{float} 
\usepackage{subfig}
\usepackage{graphicx} 

\usepackage{hyperref}

\usepackage{makecell}

\hypersetup{hidelinks,
	colorlinks=true,
	allcolors=black,
	pdfstartview=Fit,
	breaklinks=true}

\title{From Volatility to Variance: A Skew-Enhanced SABR Model and Its Empirical Study in the Chinese Financial Options Market}

\author{
 Wenxuan Zhang \\
  SKLMS, ICMSEC, NCMIS, Academy of Mathematics and Systems Science, \\ Chinese Academy of Sciences, Beijing 100190, China \\
  School of Mathematical Sciences, University of Chinese Academy of Sciences, Beijing 100049, China \\
  \texttt{zhangwenxuan17@mails.ucas.ac.cn} \\
   \And
 Zhouchi Lin \\
  International Digital Economy Academy, Shenzhen, China \\  
  \texttt{linzhouchi@idea.edu.cn} 
  \\
  \And
Benzhuo Lu$^*$  \\
   SKLMS, ICMSEC, NCMIS, Academy of Mathematics and Systems Science, \\ Chinese Academy of Sciences, Beijing 100190, China \\
  School of Mathematical Sciences, University of Chinese Academy of Sciences, Beijing 100049, China \\
  \texttt{bzlu@lsec.cc.ac.cn} \\
}

\begin{document}
\maketitle
\begin{abstract}
Accurately characterizing the implied volatility curves is a central challenge in option pricing and risk management. The classical SABR model by Hagan et al. has been widely adopted in practice due to its well-defined stochastic volatility structure and its tractable closed-form approximation for Black implied volatility. However, under complex market conditions, its fitting accuracy for implied volatility curves remains limited. To address this issue, this paper proposes an extended model within the SABR framework, referred to as skew-SABR.

Specifically, the proposed approach introduces an extension to the stochastic dynamics of the underlying asset price and its variance process, under which a corresponding Black implied volatility expression is derived. By further simplifying and reorganizing the resulting formula, the implied volatility can be expressed in a form that explicitly incorporates a skew parameter, thereby enabling a direct characterization of the asymmetry in the implied volatility curve. The resulting expression preserves the structural simplicity of the Hagan-SABR formula, while significantly enhancing the model's flexibility in capturing complex volatility smile patterns.

From a theoretical perspective, the paper provides a systematic analysis of the model specification and the financial interpretation of its parameters. From an empirical perspective, a comprehensive comparison is conducted using data from the Chinese options market over the period 2018--2025. The skew-SABR model is evaluated against the classical Hagan-SABR model, the SVI parameterization, polynomial fitting, and spline-based methods. Numerical results show that, across different market regimes and a wide range of implied volatility curve shapes, the skew-SABR model consistently achieves high and stable fitting accuracy.

Overall, the skew-SABR model preserves the theoretical structure and computational tractability of the SABR framework, while significantly improving the ability to capture asymmetric features of implied volatility curves. It thus provides a modeling approach with both solid theoretical foundations and strong practical relevance for implied volatility modeling.
\end{abstract}

\section{Introduction}

The implied volatility curve is one of the central objects in derivatives pricing and risk management, as its shape reflects the market’s aggregated expectations of future uncertainty in the underlying asset. Since the introduction of the Black--Scholes--Merton (BSM) model \cite{black1973pricing}, modern option pricing theory has been firmly established. However, due to its assumption of constant volatility 
that does not vary with strike price and time to maturity, the model fails to explain the empirically observed volatility smile phenomenon \cite{rubinstein1994implied,avellaneda1997calibrating}. This limitation has motivated extensive research on more flexible volatility modeling approaches.

To overcome the shortcomings of the constant volatility assumption, early studies introduced the local volatility framework, in which volatility is modeled as a function of the underlying asset price and time, allowing an exact fit to the market implied volatility surface in theory. The implied local volatility equation derived by Dupire provides a theoretical basis for this class of models \cite{dupire1994pricing,buraschi2001price}, followed by extensive studies on numerical implementation and stability analysis \cite{derman1994riding,andreasen2011volatility}. Despite their strong cross-sectional fitting performance, local volatility models exhibit limitations in capturing volatility dynamics and often lead to dynamic inconsistencies in hedging and risk management. Moreover, when the implied volatility surface exhibits non-monotonic smile patterns, such models may lead to non-stationarities in their dynamic evolution \cite{andersen2001extended}.

Stochastic volatility models address these limitations by introducing an additional stochastic process for volatility. The Heston model provides a closed-form characteristic function, making option pricing under stochastic volatility computationally tractable \cite{heston1993closed}. It naturally generates volatility smile patterns and has been widely applied in equity and foreign exchange markets. Hagan et al.\ proposed the SABR (Stochastic Alpha Beta Rho) model \cite{hagan2002managing}, hereafter referred to as the Hagan-SABR model. 
Through the parameters $\alpha$ (initial volatility), $\beta$ (elasticity), $\rho$ (correlation), and $\nu$ (volatility of volatility), the model characterizes the shape of the implied volatility curve and provides an approximate implied volatility formula, achieving a balance between computational efficiency and fitting accuracy. The SABR model has become a standard tool in derivatives markets.

However, the classical  Hagan-SABR formula may exhibit noticeable errors for extreme strikes or highly asymmetric volatility structures \cite{antonov2012advanced}. Existing improvements mainly follow two directions. The first introduces higher-order correction terms within the asymptotic expansion framework. For example, Paulot derived higher-order asymptotic corrections \cite{paulot2015asymptotic}, while Antonov and Spector provided a refined analytical approximation in the zero-correlation case \cite{antonov2012advanced}. In addition, Labordère et al.\ further improved approximation accuracy via asymptotic analysis \cite{henry2008analysis}. The second direction modifies the model structure, leading to various SABR extensions such as mean-reverting SABR and normal SABR, which are designed to accommodate negative interest rate environments and different market features \cite{perederiy2025mean,henry2005general,henry2008analysis}.

Although the above methods have achieved improvements in fitting accuracy, the corresponding formulas are typically structurally complex and lengthy in expression, which to some extent compromises the simplicity and practicality of the original Hagan–SABR formula. Therefore, there remains
research space for structural improvements to the SABR framework.

In contrast to models based on underlying dynamics, parametric approaches directly specify functional forms for implied volatility or total variance. The SVI (Stochastic Volatility Inspired) model, originally developed by Merrill Lynch and later formalized by Gatheral, provides a parsimonious parameterization of the implied variance curve and can be made arbitrage-free under suitable constraints \cite{gatheral2004parsimonious,gatheral2014arbitrage}. Subsequent work further established arbitrage-free conditions and parameter constraints \cite{gatheral2011volatility,gatheral2014arbitrage}. Although widely used in practice, SVI relies on a single curvature parameter to control the overall shape. When the implied volatility curve exhibits higher-order asymmetry (e.g., different curvature in the left and right tails), SVI may struggle to capture both sides simultaneously, as discussed in Section~\ref{skew-sec4}. In addition, polynomial and spline-based methods can improve fitting accuracy but may introduce oscillatory behavior and lack clear financial interpretation \cite{coleman2001reconstructing,fengler2005semiparametric}.

Other modeling approaches have also been proposed. For instance, to explain the steepening of implied volatility wings, Merton introduced the jump-diffusion model \cite{merton1976option}, and Kou later proposed the double-exponential jump-diffusion model \cite{kou2002jump}. While such models can capture extreme tail behavior, they involve complex parameter estimation and can be computationally demanding in higher dimensions. More recently, machine learning techniques have been applied to volatility surface modeling. Horvath et al.\ employed deep neural networks for option pricing and volatility surface fitting \cite{horvath2021deep}. Although these methods often achieve high empirical accuracy, they face challenges in enforcing no-arbitrage conditions, ensuring financial interpretability, and achieving robust cross-market generalization.

In summary, existing approaches to implied volatility modeling can be broadly classified into three categories.

The first category consists of structural models based on the risk-neutral dynamics of the underlying asset, including local volatility and stochastic volatility models. These models provide clear economic interpretations and dynamic consistency. However, they typically rely on strong structural assumptions and often require asymptotic expansions or numerical methods to obtain implied volatility approximations. Their flexibility may be limited when the volatility surface exhibits complex curvature or asymmetry.

The second category includes parametric and nonparametric approaches that directly model implied volatility or total variance in the cross section. Parametric methods, such as SVI or polynomial fitting, use a finite number of parameters, while nonparametric (or semiparametric) methods, such as spline interpolation, adjust model flexibility through the number of knots. These approaches are relatively simple and stable in calibration and offer strong cross-sectional flexibility. However, they generally lack an explicit stochastic process foundation, limiting the financial interpretability of parameters and their dynamic consistency across maturities.

The third category consists of data-driven approaches based on statistical learning or machine learning. These methods leverage high-dimensional function approximation and often achieve strong empirical performance. However, their structure is largely data-driven, with limited theoretical grounding and financial interpretation, and no-arbitrage conditions typically need to be imposed separately.

Against this background, this paper proposes an improved model within the SABR framework, referred to as skew-SABR. Specifically, we reformulate the joint dynamics of the underlying asset price and its volatility process and derive the corresponding Black implied volatility expression under the new specification. By further simplifying the resulting formula, the implied volatility can be expressed in a form that explicitly incorporates a skew parameter, enabling a direct characterization of asymmetry in the implied volatility curve. This representation preserves the structural simplicity of the Hagan-SABR formula while enhancing its flexibility in capturing complex volatility patterns. As a result, the proposed model provides improved fitting performance across different market conditions.

The main contributions of this paper are summarized as follows:

(1) A new stochastic process model is constructed within the SABR framework, and its stochastic differential equation under the risk-neutral measure is specified. Based on this framework, a corresponding Black implied volatility formula is derived, leading to the skew-SABR model.

(2) The derived implied volatility expression is systematically simplified to obtain a tractable approximation suitable for practical implementation. The relationship between the new model parameters and those in the original Hagan-SABR model is analyzed, and the financial interpretation of each parameter in capturing skewness and curvature is discussed.

(3) Under a unified empirical framework, the skew-SABR model is systematically compared with the classical Hagan-SABR model, the SVI parameterization, polynomial methods, and spline-based approaches. The analysis focuses on scenarios with asymmetric skew and higher-order structural features. Numerical results show that the skew-SABR model achieves higher and more stable fitting accuracy for complex implied volatility curves.

The remainder of this paper is organized as follows. Section~\ref{skew-sec2} reviews the classical Hagan-SABR model and its implied volatility approximation, and discusses its limitations. Section~\ref{skew-sec3} presents the proposed skew-SABR model, including its stochastic dynamics and the derivation of the implied volatility formula, along with an analysis of parameter interpretations. Section~\ref{skew-sec4} provides empirical comparisons under a unified framework. Section~\ref{skew-sec5} concludes and outlines directions for future research.

\section{The Classical Hagan--SABR Model and Its Fitting Performance}

\label{skew-sec2}

The SABR model was originally proposed by Hagan et al. \cite{hagan2002managing}. Its primary motivation is to capture the implied volatility smile widely observed in financial markets.
A single Brownian motion is insufficient to reproduce this stylized fact. Therefore, an additional stochastic process is introduced to model the volatility itself, leading to a stochastic volatility framework. Under the risk-neutral measure, the Hagan--SABR model assumes that the forward price of the underlying asset and its volatility process satisfy the following system of stochastic differential equations:
\begin{equation}
\left\{  
\begin{aligned} 
& \quad d\hat{F} = \hat{\alpha}\hat{F}^{\beta}dW_1, \quad \hat{F}(0) = f, \\ 
&\quad d\hat{\alpha} = \nu \hat{\alpha}dW_2, \quad \hat{\alpha}(0) = \alpha, \\ 
&\quad dW_1 dW_2 = \rho dt.
\end{aligned}
\right.   \label{skew-1}
\end{equation}
Here, $\hat{F}_t$ denotes the forward price process with maturity $T$, and $\hat{\alpha}_t$ represents the stochastic volatility process. 
$W_1(t)$ and $W_2(t)$ are two correlated standard Brownian motions with correlation coefficient $\rho$. 
SABR stands for ``Stochastic Alpha Beta Rho,'' and each parameter has a clear financial interpretation: 
$\alpha$ represents the initial volatility level; 
$\beta \in [0,1]$ is the elasticity parameter controlling the dependence of the diffusion term on the price level; 
$\rho$ captures the correlation between the underlying price and its volatility; 
$\nu$ is the volatility of volatility (vol-of-vol), measuring the randomness of the volatility process itself. 
These parameters jointly characterize the structural features of the implied volatility surface.

Within this stochastic volatility framework, Hagan et al. employed singular perturbation techniques to derive an asymptotic expansion for the Black implied volatility under the SABR model. 
Substituting the resulting implied volatility into the Black pricing formula yields an approximate option price, enabling the characterization of market option prices. 
They provide a more general formulation of the model.
Specifically, consider the following generalized system:
\begin{equation}
\left\{  
\begin{aligned} 
& \quad d\hat{F} = \hat{\alpha}C(\hat{F})dW_1, \quad \hat{F}(0) = f, \\ 
&\quad d\hat{\alpha} = \nu \hat{\alpha}dW_2, \quad \hat{\alpha}(0) = \alpha, \\ 
&\quad dW_1 dW_2 = \rho dt.
\end{aligned}
\right.  \label{skew-2}
\end{equation}
where $C(\hat{F}_t)$ is a deterministic function of the forward price. 
In the classical Hagan--SABR model (Eq.~\eqref{skew-1}), we have $C(F_t) = F_t^{\beta}$.

Under this general framework, the Black implied volatility $\sigma_B(K)$ admits the asymptotic expansion
\begin{equation*}
    \sigma_B(K) = I_0 \cdot (1 + I_1 \tau + O(\tau^2)),
\end{equation*}
where $I_0$ and $I_1$ are functions of the model parameters and the strike price $K$, and $O(\tau^2)$ denotes higher-order terms in the time to maturity $\tau$. 
In practice, these higher-order terms are typically neglected, yielding a first-order approximation denoted by $\sigma_B(\alpha, \beta, \rho, \nu; K, \tau)$. 
After truncating higher-order terms, the implied volatility corresponding to Eq.~\eqref{skew-2} can be expressed as
\begin{equation}
    \sigma_B(\alpha, \beta, \rho, \nu; K, \tau) = I_0 (\alpha, \beta, \rho, \nu; K, \tau)\cdot \big[1 + I_1 (\alpha, \beta, \rho, \nu; K, \tau) \cdot \tau \big]. \label{skew-3}
\end{equation}
where the explicit expressions of $I_0$ and $I_1$ are given by
\begin{subequations}
\begin{align}
&I_0(\alpha, \beta, \rho, \nu; K, \tau) = \frac{\alpha \log \frac{f}{K}}{\int_K^f \frac{dx}{C(x)}} \cdot  \bigg (\frac{\zeta}{\chi(\zeta)}\bigg ),  \label{skew-4a}  \\
I_1(\alpha, \beta, \rho, \nu; K, \tau) =& \frac{2\gamma_2-\gamma_1^2 + 1/f_{ave}^2}{24} \alpha^2 C^2(f_{ave}) + \frac{1}{4}\rho \nu \alpha \gamma_1 C(f_{ave}) + \frac{2-3\rho^2}{24}\nu^2,  \\
f_{ave} = \sqrt{fK}, \quad &\gamma_1 = \frac{C'(f_{ave})}{C(f_{ave})}, \quad \gamma_2 = \frac{C''(f_{ave})}{C(f_{ave})}, \quad \tau = T-t, \\ \zeta =&  \frac{\nu}{\alpha}\int_K^f \frac{dx}{C(x)}, \quad \chi(z) = \log \bigg (\frac{\sqrt{1-2\rho z + z^2} - \rho + z}{1 - \rho}\bigg ). \label{skew-4d}
\end{align}
\end{subequations}

For the classical Hagan-SABR model in Eq.~(\ref{skew-1}), we have $C(f) = f^{\beta}$. Therefore, the integral term in Eq.~(\ref{skew-4a}) can be written as
\begin{align*}
\int_K^f \frac{dx}{C(x)} = \int_K^f \frac{dx}{x^{\beta}} &=
\begin{cases}
\frac{1}{1-\beta}(f^{1-\beta}-K^{1-\beta}) & \text{if}\ 0 \leq \beta < 1,\\
\log \frac{f}{K} & \text{if}\ \beta = 1.
\end{cases}
\end{align*}

To further obtain an explicit approximation, the following logarithmic expansions can be used:
\begin{align*}
    f - K &= \sqrt{fK}\log \frac{f}{K} \Big(1+\frac{1}{24}\log^2\frac{f}{K}+\frac{1}{24}\log^4\frac{f}{K} + \cdots \Big), \\
    f^{1-\beta} - K^{1-\beta} &= (1-\beta)(fK)^{(1-\beta)/2}\log \frac{f}{K} \Big(1 + \frac{(1-\beta)^2}{24}\log^2\frac{f}{K} + \frac{(1-\beta)^4}{1920}\log^4 \frac{f}{K} + \cdots \Big).
\end{align*}

Substituting the above results into Eqs.~(\ref{skew-4a})–(\ref{skew-4d}), one can derive the explicit forms of $I_0$ and $I_1$. Further substituting them into Eq.~(\ref{skew-3}) yields the approximate Black implied volatility under the Hagan-SABR model, denoted by $\sigma_B^{H\text{-}SABR}$, where the superscript indicates the model name.

\begin{equation} 
\begin{split}
\sigma_B^{H\text{-}SABR} = &\frac{\alpha}{(fK)^{(1-\beta)/2}}\frac{1}{1 + \frac{(1-\beta)^2}{24}\log^2\frac{f}{K} + \frac{(1-\beta)^4}{1920}\log^4 \frac{f}{K}} \cdot \frac{\zeta}{\chi(\zeta)} \cdot \\ 
& \bigg[ 1 + \Big(\frac{(1-\beta)^2\alpha^2}{24(fK)^{1-\beta}} + \frac{\rho \alpha \nu \beta}{4(fK)^{(1-\beta)/2}} + \frac{2-3\rho^2}{24}\nu^2 \Big)\tau \bigg]
\end{split} 
\label{skew-8}
\end{equation}

Eq.~(\ref{skew-8}) is the approximate implied volatility formula originally derived by Hagan et al. in their SABR model. In practical option pricing, one can replace the constant volatility $\sigma$ in the Black–Scholes formula with $\sigma_B^{H\text{-}SABR}$ to obtain an approximate market-consistent option price.

As shown in Eq.~(\ref{skew-8}), the Hagan-SABR model is characterized by four parameters: $\alpha$, $\beta$, $\rho$, and $\nu$. Given the forward price $f$ and the time to maturity $\tau$, specifying these parameters allows the Black implied volatility to be computed for any strike $K$ via Eq.~(\ref{skew-8}), thereby producing a complete implied volatility curve.

These four parameters influence the shape of the implied volatility curve from different perspectives and thus play distinct structural roles in model calibration. To illustrate the impact of each parameter on the shape of the implied volatility surface, we fix a set of baseline parameters
$\alpha = 0.25,\ \beta = 0.9,\ \rho = -0.2,\ \nu = 1$, and adopt a ceteris paribus approach, varying one parameter at a time while keeping the others fixed.

For the market variables such as the forward price, time to maturity, and strike prices, we use actual market data as reference. Specifically, we select the CSI 1000 equity index option with maturity in June 2025 (contract code: MO202506) at 9:50:05 on April 17, 2025 as the baseline. The corresponding forward price is $F = 5685.6$, the time to maturity is $\tau = 0.176$, and the strike range is set to $[3900,\,7400]$, with strikes sampled at intervals of 100.

\begin{figure}[htbp]
    \centering
\includegraphics[width=0.99\textwidth]{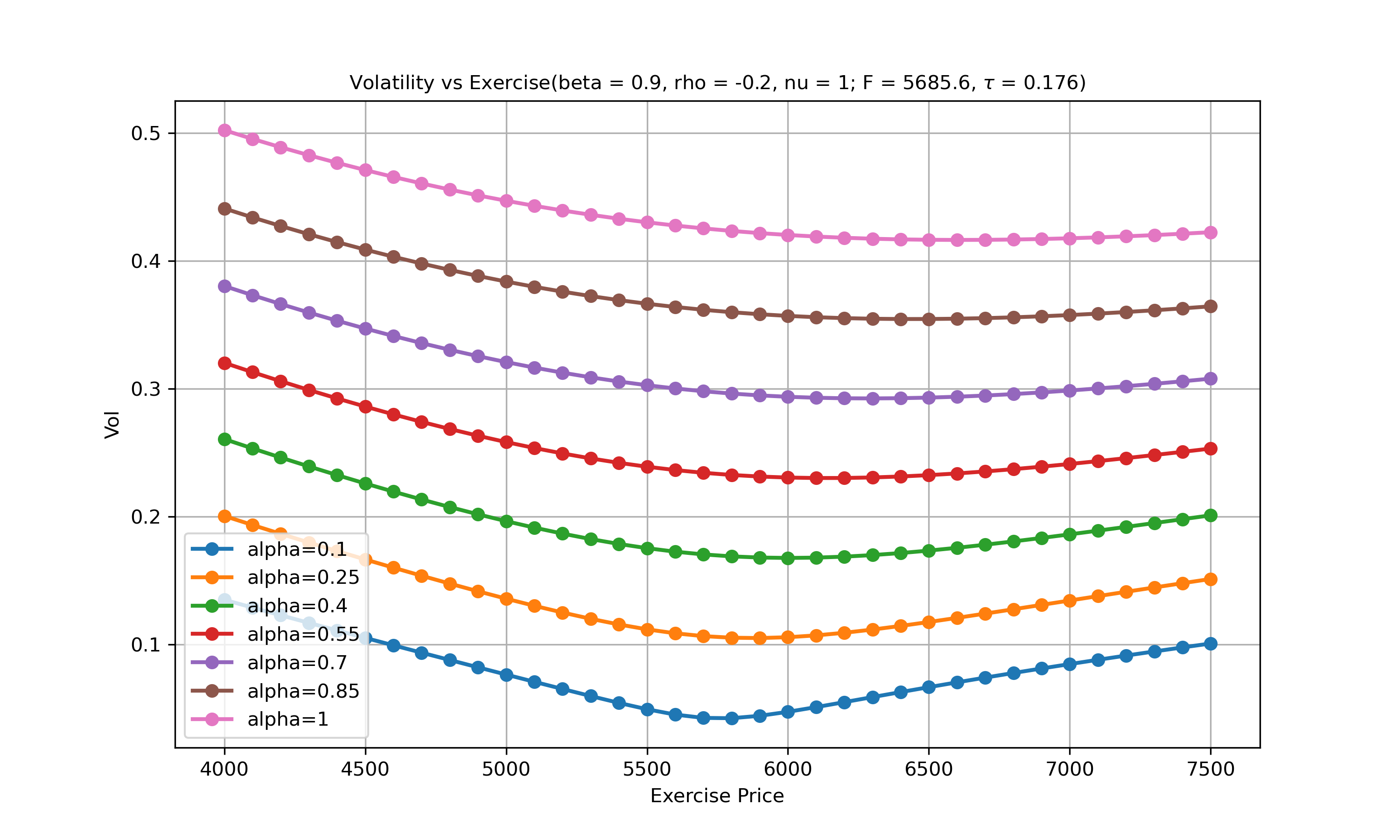}\\
\caption{Implied volatility curves generated by the Hagan-SABR model under different values of $\alpha$} \label{skew-fig3}
\end{figure}

First, the parameter $\alpha$ represents the initial volatility level. As shown in Fig.~\ref{skew-fig3}, $\alpha$ primarily controls the overall level of the implied volatility curve, corresponding to the general level of volatility observed in the market.

\begin{figure}[htbp]
    \centering
\includegraphics[width=0.99\textwidth]{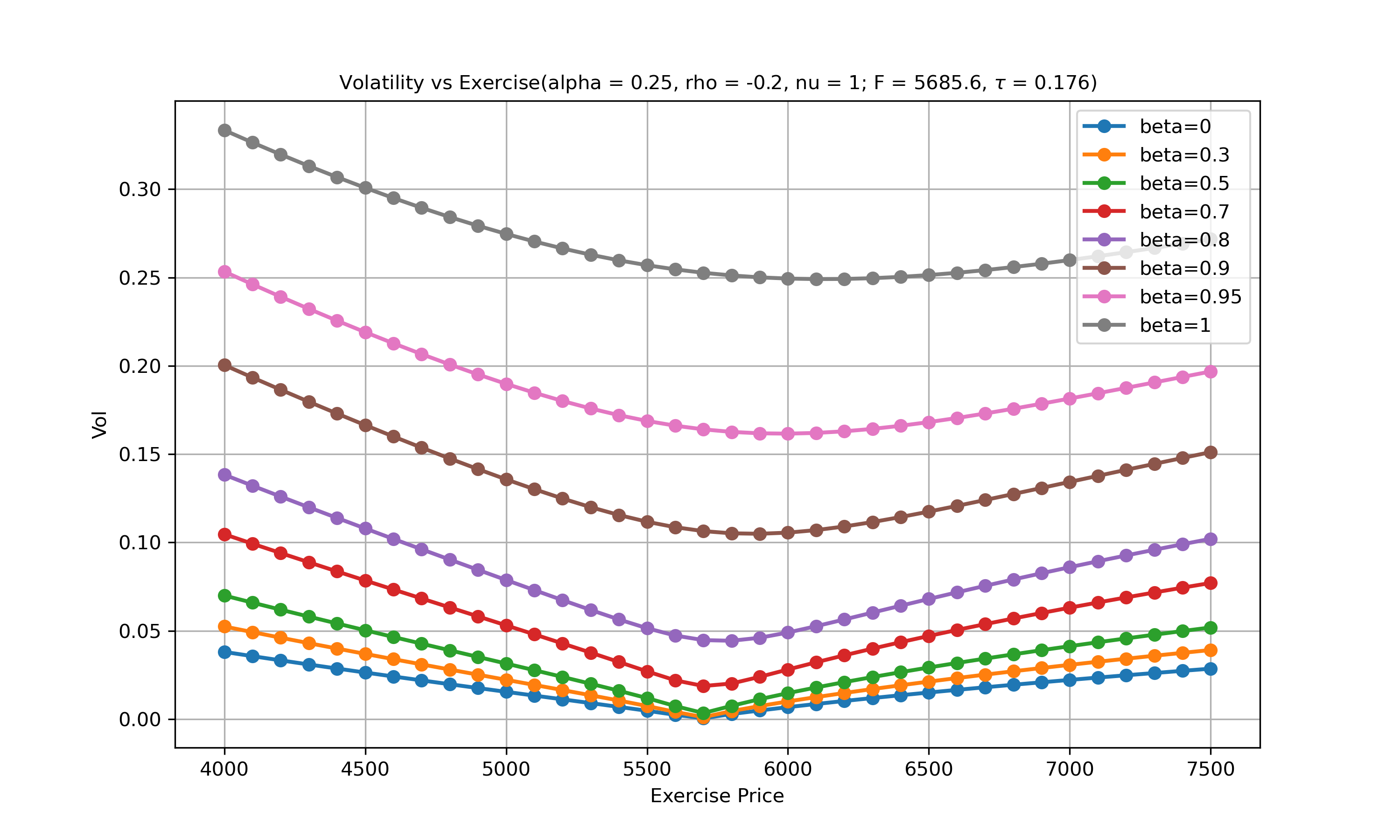}\\
\caption{Implied volatility curves generated by the Hagan-SABR model under different values of $\beta$} \label{skew-fig2}
\end{figure}

The parameter $\beta$, known as the constant elasticity of variance (CEV) parameter, determines the type of stochastic process followed by the underlying asset and typically takes values in the interval $[0,1]$. When $\beta=0$, the model corresponds to a normal (Bachelier-type) structure; when $\beta=1$, it corresponds to a lognormal (Black-type)  structure. By adjusting $\beta$, the model is able to capture different distribution characteristics of the underlying asset. Fig.~\ref{skew-fig2} illustrates the effect of different values of $\beta$ on the shape of the curve.

\begin{figure}[htbp]
    \centering
\includegraphics[width=0.99\textwidth]{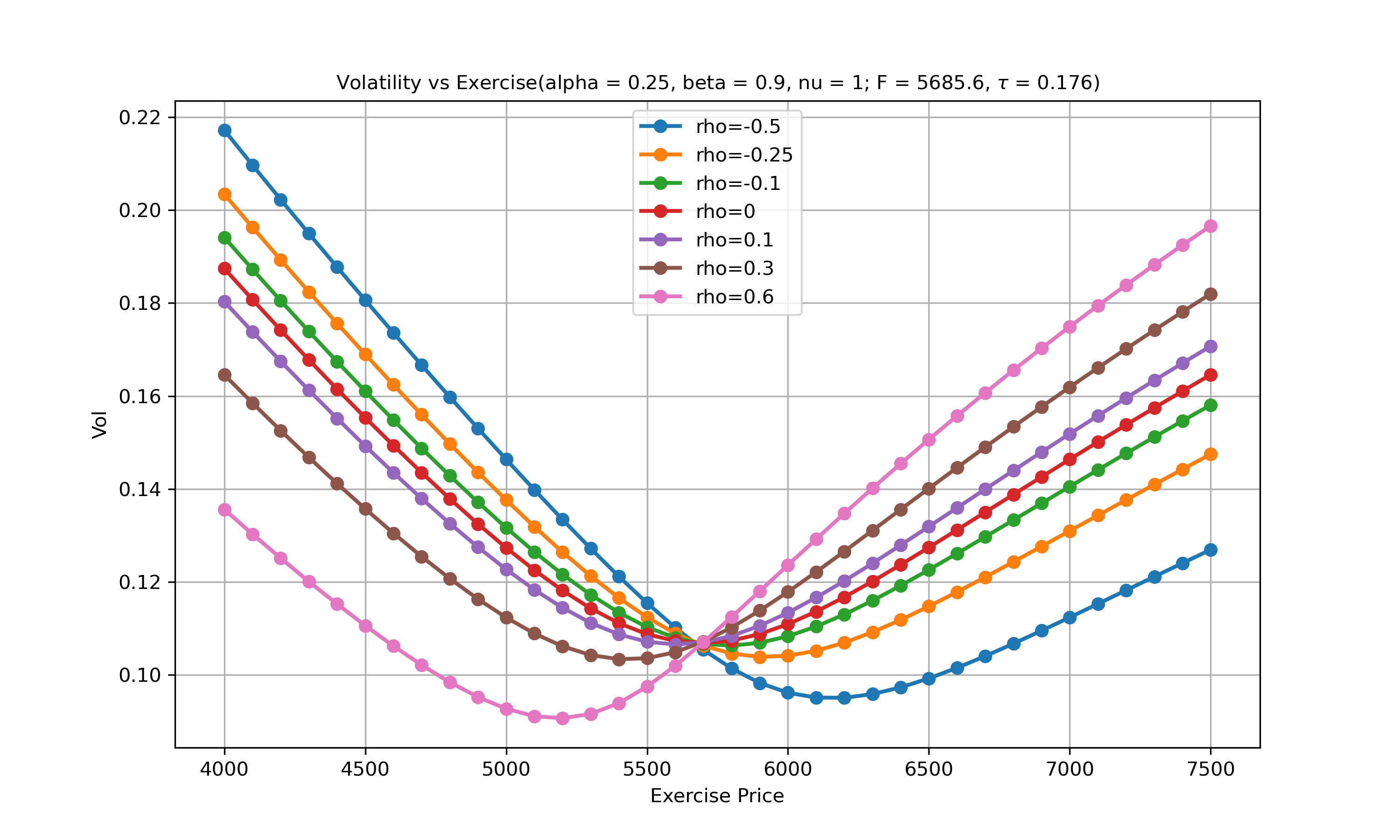}\\
\caption{Implied volatility curves generated by the Hagan-SABR model under different values of $\rho$} \label{skew-fig4}
\end{figure}

The parameter $\rho$ denotes the correlation between the underlying asset price and its volatility process, with values in the range $[-1,1]$. Both the magnitude and sign of $\rho$ directly affect the skewness of the implied volatility curve. As illustrated in Fig.~\ref{skew-fig4}, variations in $\rho$ lead to a leftward or rightward tilt of the curve, reflecting different market expectations and risk preferences.

To further clarify this effect, we introduce the concept of skewness, which is a key characteristic of the implied volatility curve and plays an important role in the subsequent model development.
The skew of the implied volatility curve refers to its overall slope with respect to the strike price, reflecting the relative pricing differences across different strikes. In essence, it captures the market's asymmetric pricing of upside and downside risks of the underlying asset. For example, when $\rho < 0$, a decrease in the underlying asset price is typically associated with an increase in volatility, resulting in a negatively skewed implied volatility curve. In this case, low-strike options (OTM puts) exhibit higher implied volatilities than high-strike options, indicating that investors are willing to pay a premium for downside protection. Conversely, when $\rho > 0$, the curve exhibits positive skewness, reflecting a relatively stronger pricing of upside risk.

\begin{figure}[htbp]
    \centering
\includegraphics[width=0.99\textwidth]{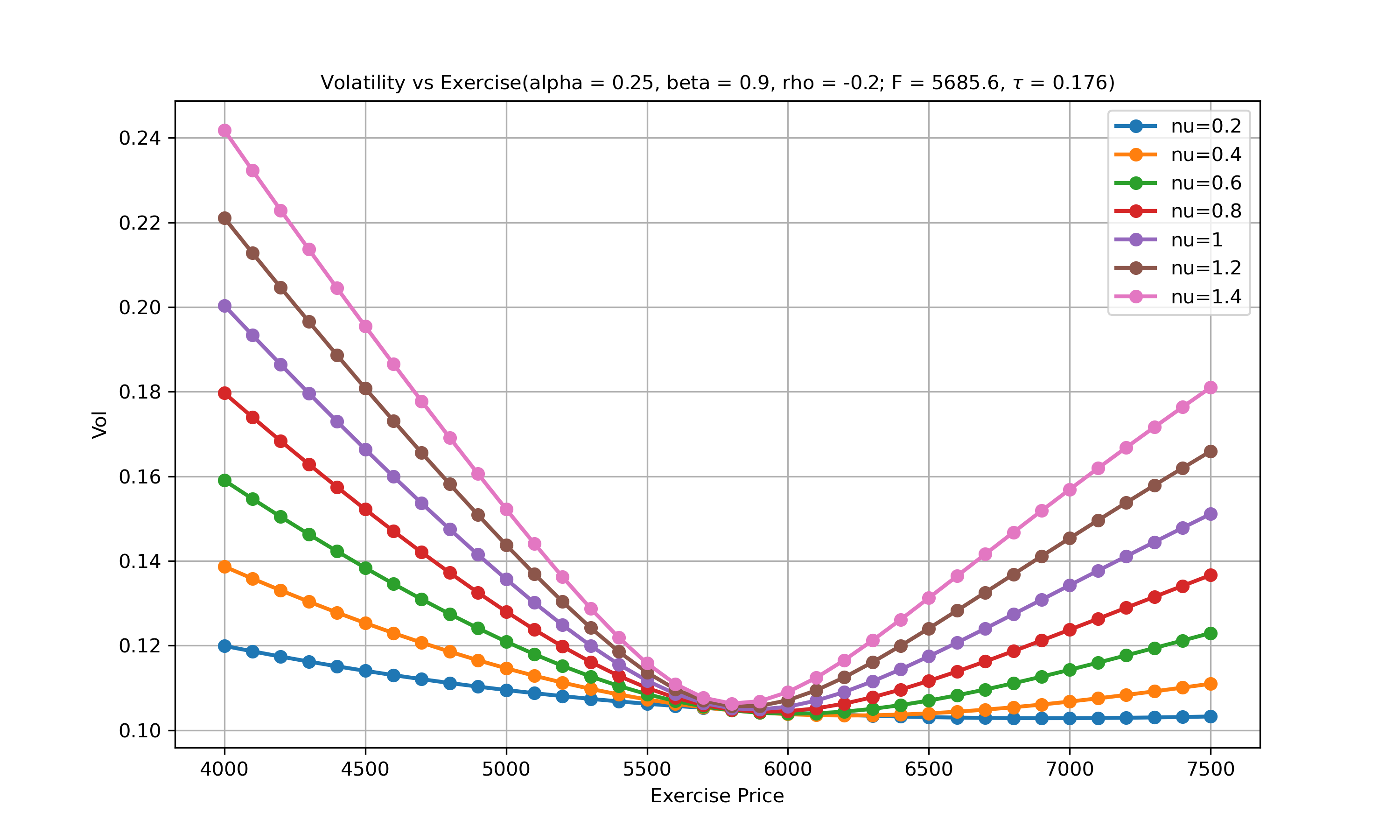}\\
\caption{Implied volatility curves generated by the Hagan-SABR model under different values of $\nu$} \label{skew-fig5}
\end{figure}

The parameter $\nu$, referred to as the volatility of volatility (vol-of-vol), measures the degree of randomness in the volatility process and takes positive values. A larger $\nu$ implies greater uncertainty in volatility, leading to a more pronounced curvature of the implied volatility curve, as shown in Fig.~\ref{skew-fig5}. Together, $\rho$ and $\nu$ determine the first-order (skewness) and second-order (curvature) characteristics of the curve. A more detailed mathematical derivation and quantitative analysis will be provided in subsequent sections.

To estimate the parameters $\alpha, \beta, \rho, \nu$, model calibration is required. 
This is typically formulated as an optimization problem that minimizes the weighted squared error between model-implied and market-implied volatilities. 
Two commonly used calibration approaches are as follows.

The first approach jointly calibrates all four parameters:
\begin{equation}
    \alpha, \beta, \rho, \nu = \arg \min \sum_K w_K |\sigma_B^{H\text{-}SABR}(K, f) - \sigma_{\text{market}}(K, f)|^2. \label{8}
\end{equation}

Here, $\sigma_B^{H\text{-}SABR}$ denotes the model-implied volatility, while $\sigma_{\text{market}}$ is the market-implied volatility obtained by
inverting the current market price through the Black-Scholes pricing formula. 
$w_K$ is a weighting function, typically assigning larger weights near the at-the-money (ATM) region to emphasize liquid strikes. 
In this paper, we adopt the following weighting scheme:
\begin{equation*}
    w_K = \exp\left[-0.25\left(\frac{K-f}{0.1f}\right)^2\right].
\end{equation*}

To simplify the calibration procedure and improve numerical stability, the second approach fixes $\beta$ and calibrates only the remaining three parameters:
\begin{equation}
    \alpha, \rho, \nu = \arg \min \sum_K w_K |\sigma_B^{H\text{-}SABR}(K, f) - \sigma_{\text{market}}(K, f)|^2. \label{9}
\end{equation}

Hagan, in work \cite{hagan2002managing}, pointed out that the simultaneous calibration of all four parameters may lead to instability in parameter estimation in practical applications. In contrast, fixing $\beta$ and calibrating only the remaining three parameters typically yields a fitting performance comparable to that of the four-parameter approach. Moreover, it is also noted that setting $\beta = 1$, which corresponds to assuming that the underlying asset follows a lognormal distribution, is a natural and commonly adopted specification.

Additional numerical checks conducted in this study (not reported in detail here) indicate that, under the four-parameter calibration framework, the estimated values of $\beta$ are generally concentrated around $1$, and the corresponding fitting performance is nearly identical to that obtained from the three-parameter calibration. Based on the above theoretical considerations and empirical observations, this paper adopts the three-parameter calibration approach with $\beta = 1$ fixed, and estimates the remaining parameters according to Eq.~(\ref{9}).

With $\beta = 1$ fixed, the dynamics of the Hagan-SABR model can be expressed as
\begin{equation}
\left\{  
\begin{aligned} 
& \quad d\hat{F} = \hat{\alpha}\hat{F}dW_1, \quad \hat{F}(0) = f, \\ 
&\quad d\hat{\alpha} = \nu \hat{\alpha}dW_2, \quad \hat{\alpha}(0) = \alpha, \\ 
&\quad dW_1dW_2 = \rho dt.
\end{aligned}
\right. \label{skew-10}
\end{equation}
For an underlying asset following the above stochastic process, the corresponding approximation of the Black implied volatility is given by
\begin{equation}
\begin{aligned}
    &\sigma^{H-SABR}_B(\alpha, \rho, \nu; K, \tau) = I_0^{H} \cdot (1 + I_1^{H} \tau) \\ &I_0^{H}(\alpha, \rho, \nu; K, \tau) = \frac{\alpha \zeta}{\chi(\zeta)}, \quad \zeta = \frac{\nu}{\alpha} \log{\frac{f}{K}} \\ &I_1^{H}(\alpha, \rho, \nu) = (\frac{1}{4}\rho \nu \alpha + \frac{2-3\rho^2}{24}\nu^2) \label{skew-12}
    \end{aligned}
\end{equation}

Here, the term $I_1^{H}$ is independent of the strike price $K$ and depends only on the model parameters $\alpha$, $\rho$, and $\nu$. Under the constraint $\beta = 1$, only these three parameters need to be calibrated. The corresponding optimization problem can be formulated as
\begin{align*}
    \alpha, \rho, \nu &= \arg \min \sum_K w_K \left|\sigma_B^{H\text{-}SABR}(K, f) - \sigma_{\text{market}}(K, f)\right|^2 \\
    &= \arg \min \sum_K w_K \left| I_0^{H}(\alpha, \rho, \nu; K, \tau) \cdot (1 + I_1^{H}(\alpha, \rho, \nu)\tau) - \sigma_{\text{market}}(K, f) \right|^2.
\end{align*}

However, although the three-parameter calibration approach with $\beta=1$ improves the temporal stability of the model parameters to some extent, the SABR model still exhibits notable deficiencies in fitting the implied volatility surface of the Chinese index options market. In particular, the fitting errors are more pronounced in the out-of-the-money (OTM) call region.

To illustrate this issue, we consider CSI 1000 index options (contract code:MO) and CSI 500 ETF (contract code:510500) options as examples, with maturities in May, June, and September 2025. At the time point 09:50:05 on April 17, 2025, the corresponding market implied volatility curves are extracted for analysis. For each strike price, the implied volatility is selected from either the call or put option with the smaller bid–ask spread. In general, put options tend to have better liquidity and tighter spreads at lower strike prices, while call options typically provide more reliable quotes at higher strike prices.

\begin{figure}[htbp]
  \subfloat[]{\includegraphics[width=0.47\textwidth]{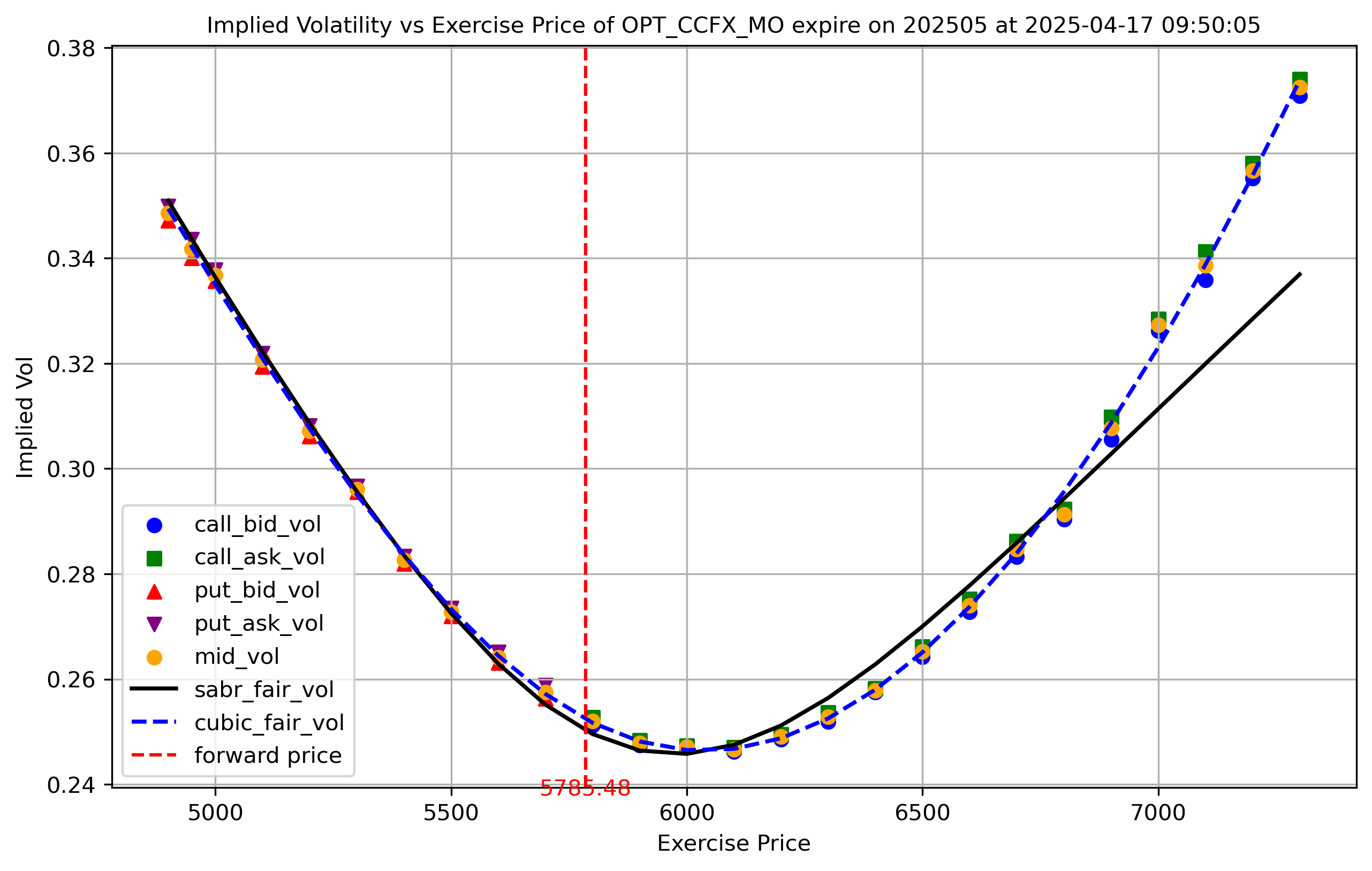}}
  \hfill	
  \subfloat[]{\includegraphics[width=0.47\textwidth]{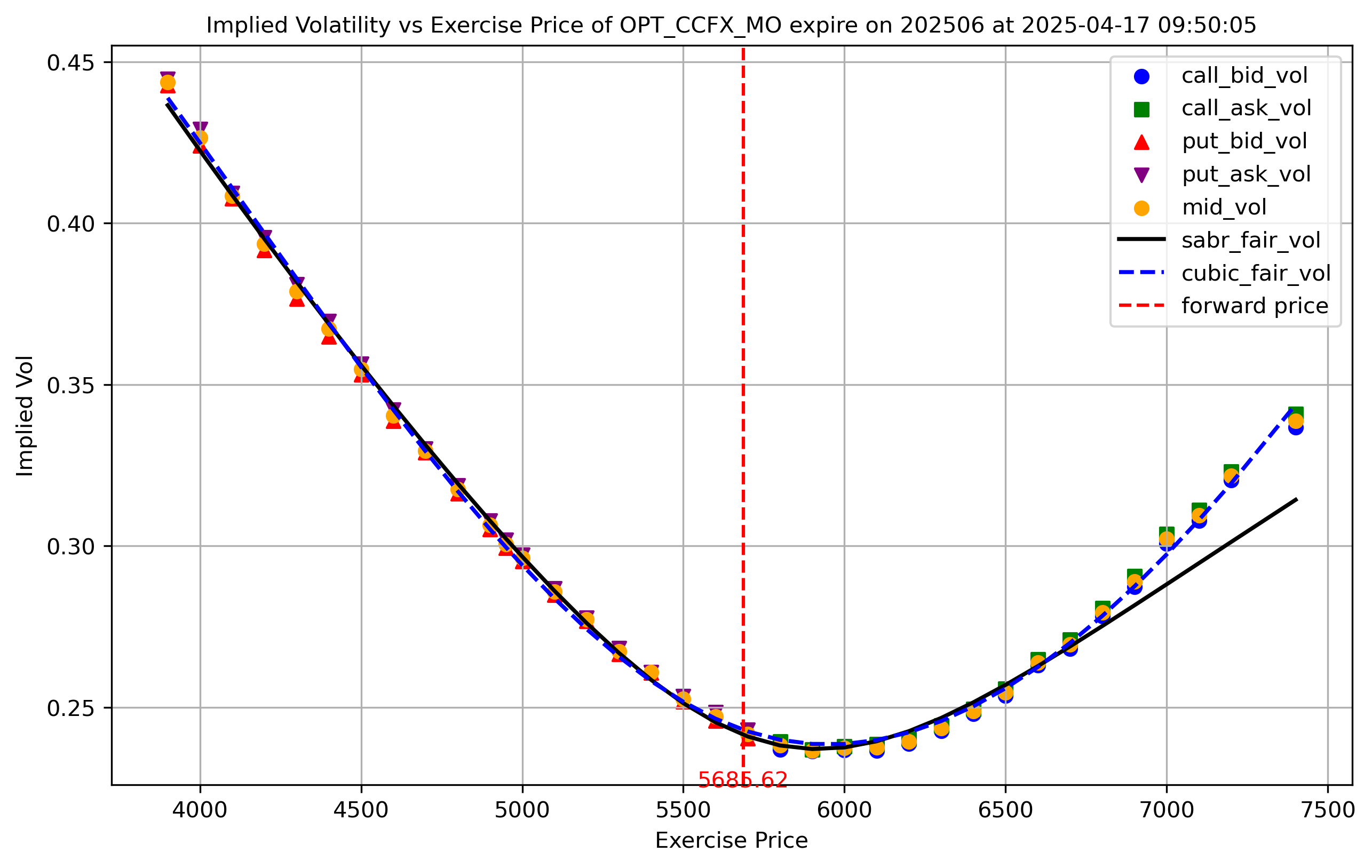}}

  \subfloat[]{\includegraphics[width=0.47\textwidth]{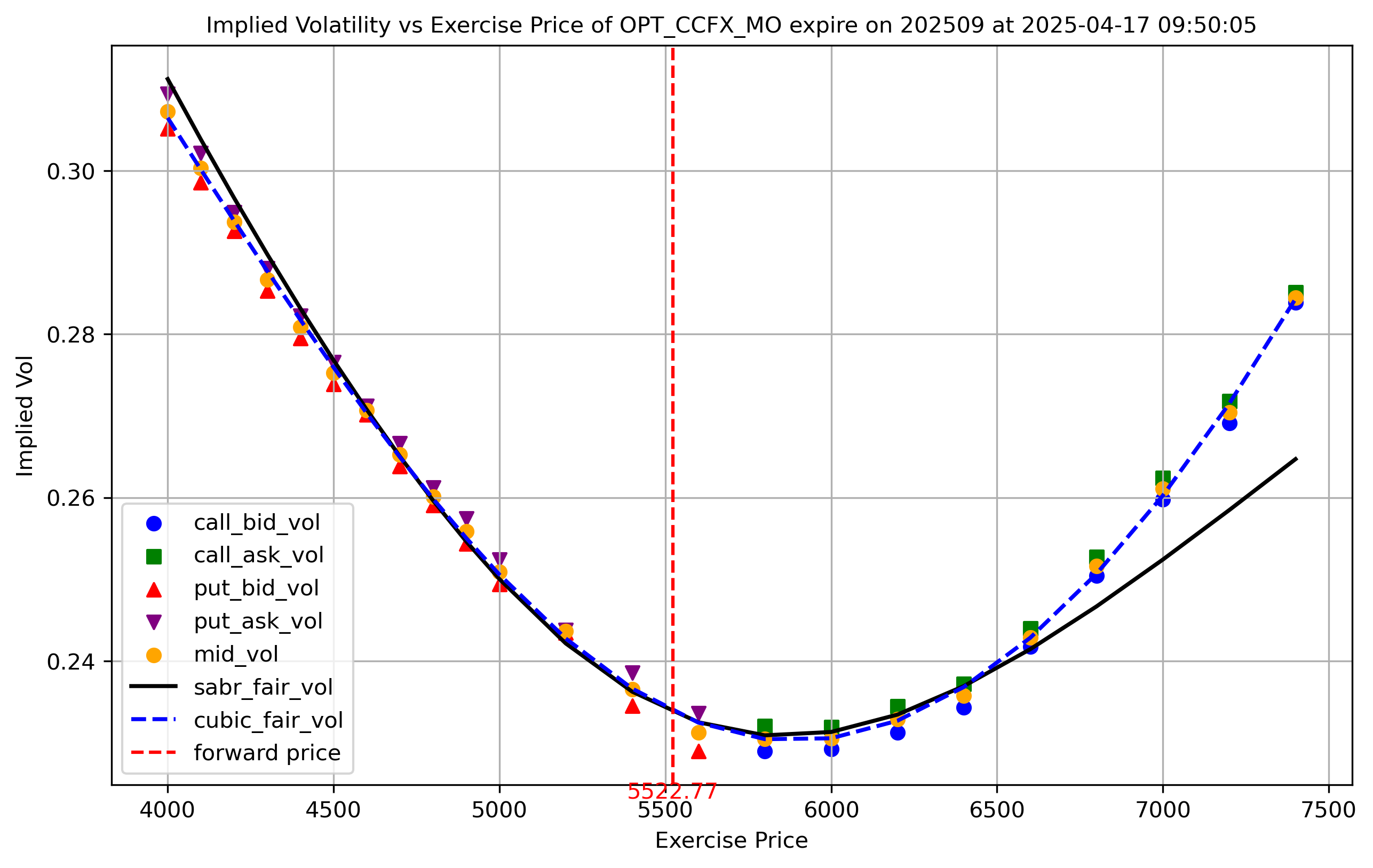}}
  \hfill	
  \subfloat[]{\includegraphics[width=0.47\textwidth]{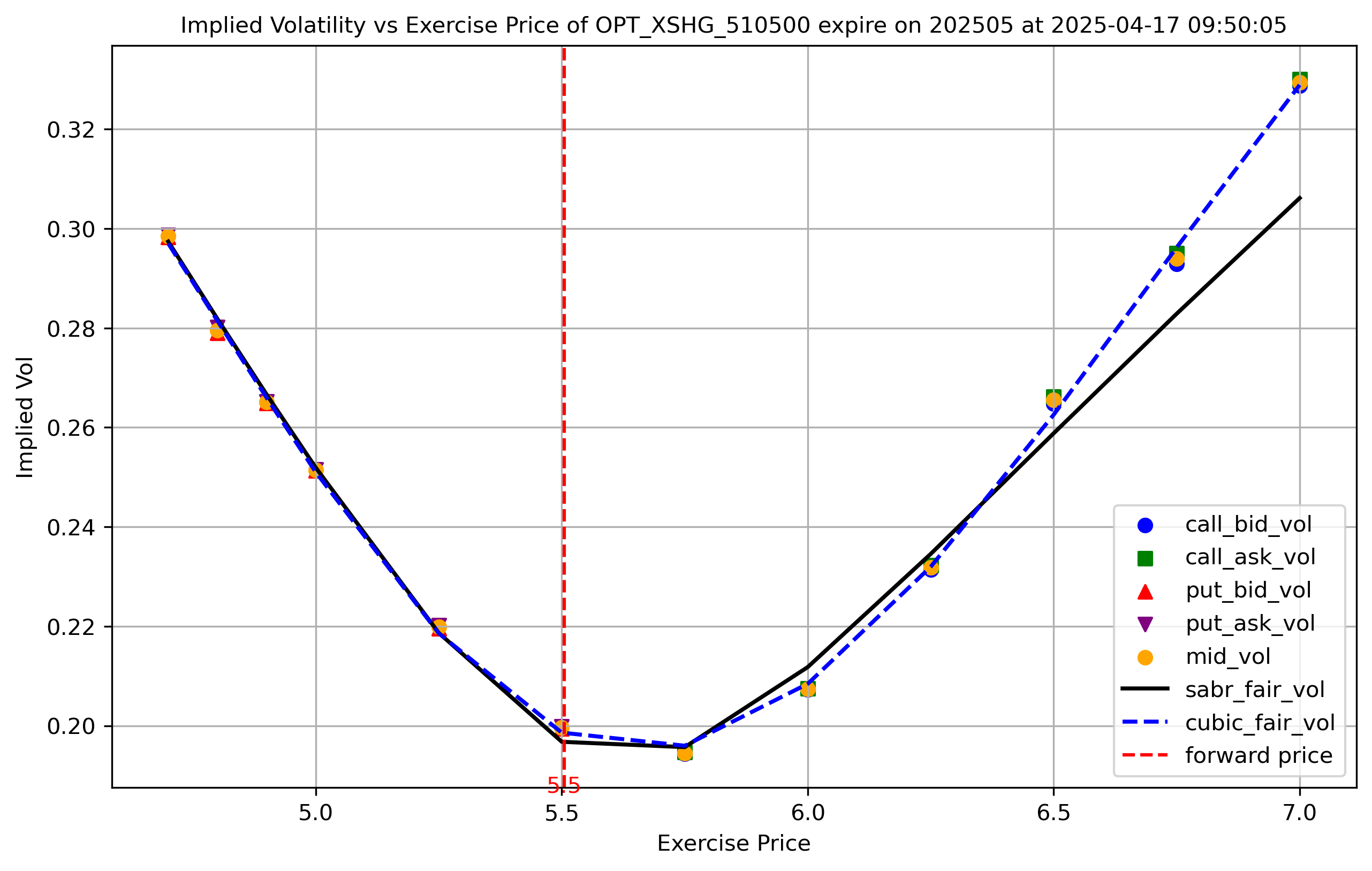}}

  \subfloat[]{\includegraphics[width=0.47\textwidth]{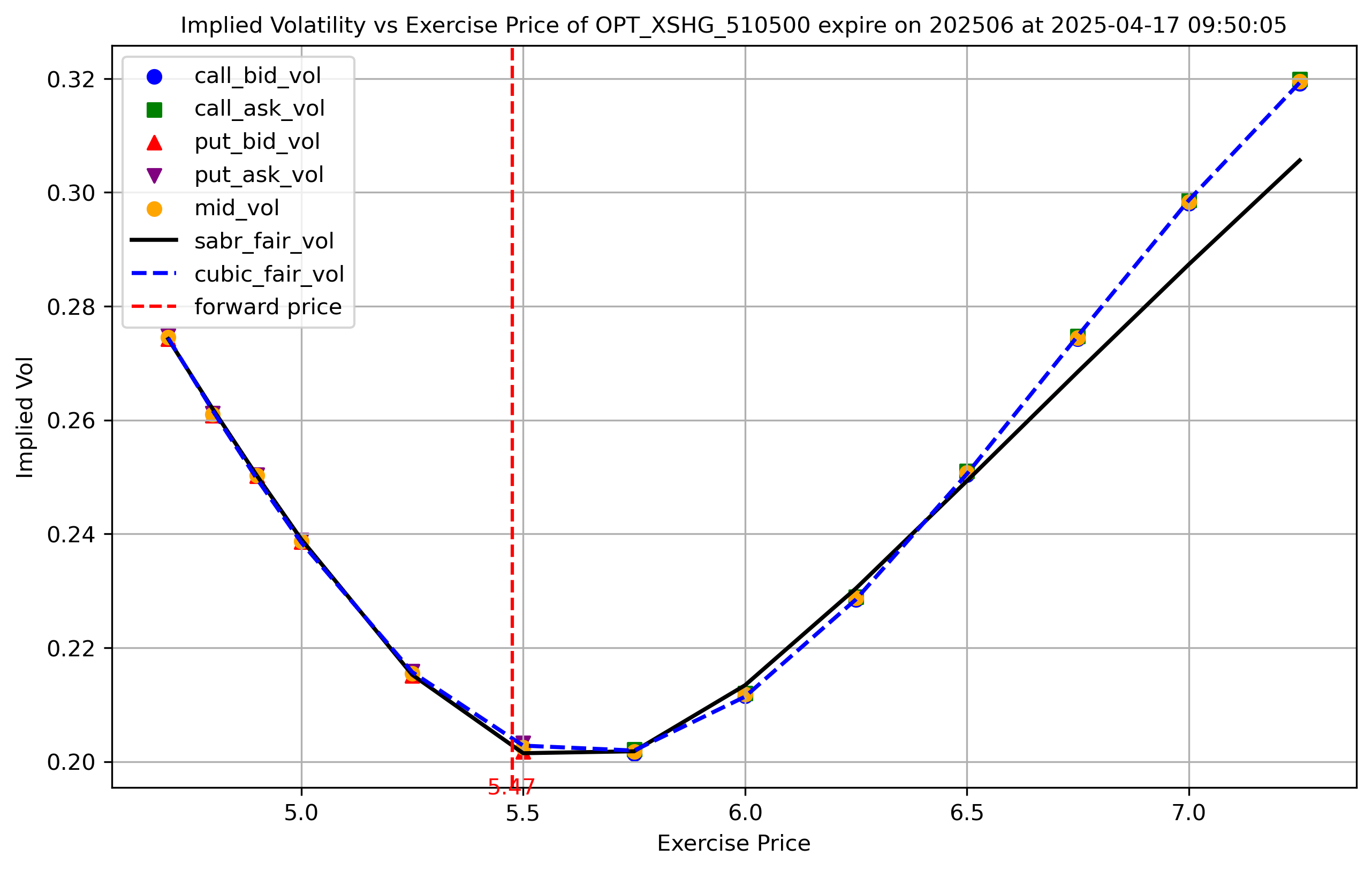}}
  \hfill	
  \subfloat[]{\includegraphics[width=0.47\textwidth]{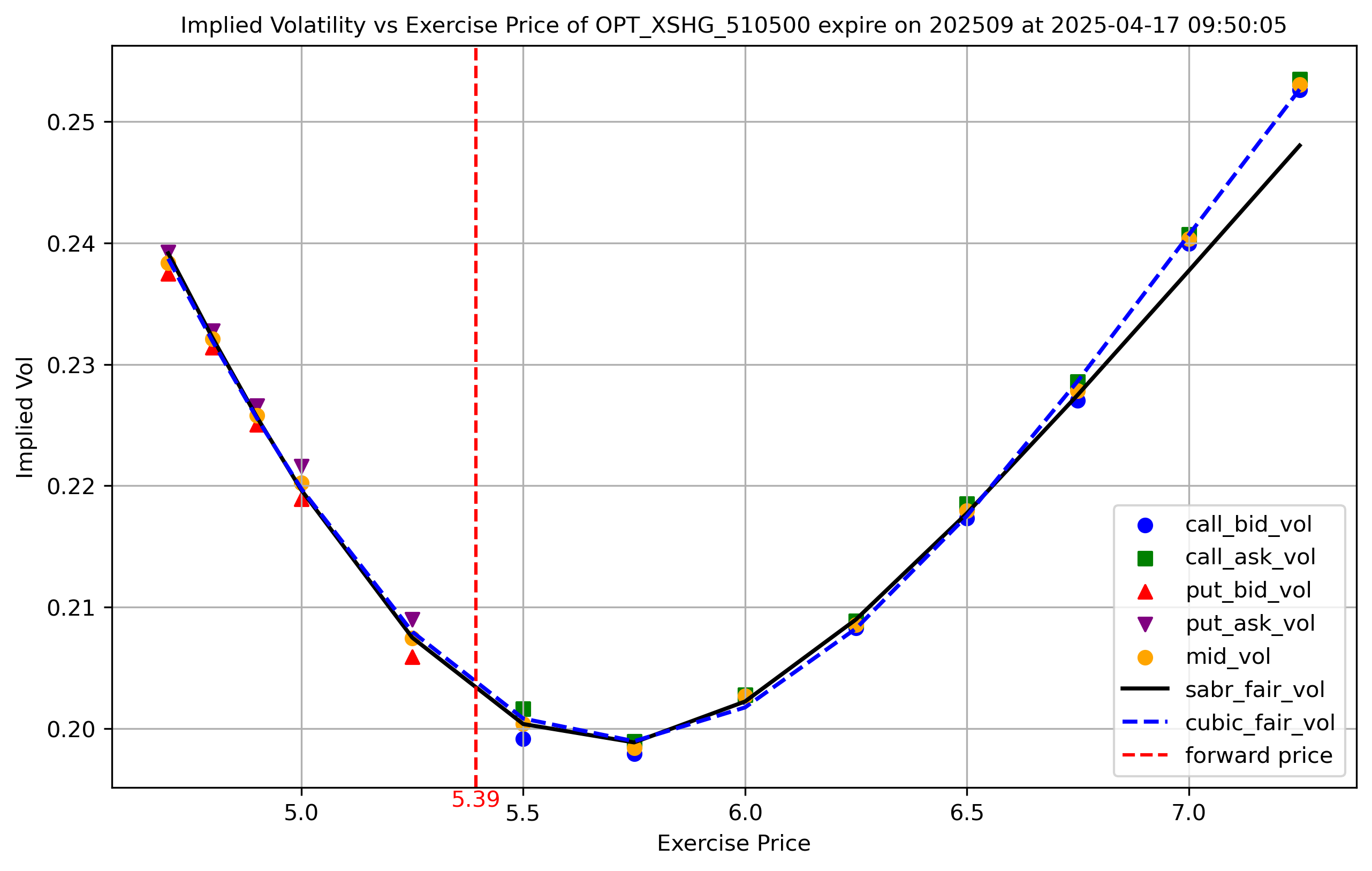}}
\caption{Implied volatility fitting results of the Hagan-SABR model across different underlying assets and maturities. (a) CSI 1000 Index Options, May 2025 Contract; (b) CSI 1000 Index Options, June 2025 Contract; (c) CSI 1000 Index Options, September 2025 Contract; (d) CSI 500 ETF option, May 2025 Contract; (e) CSI 500 ETF option, June 2025 Contract; (f) CSI 500 ETF option, September 2025 Contract.}  \label{skew-fig16}
\end{figure}

Based on these data, the SABR model is calibrated to the implied volatility curves, and the fitted results are compared with the market observations, as shown in Fig.~\ref{skew-fig16}. The blue dashed lines represent the reference curves obtained via spline interpolation (see Section~\ref{skew-sec4} for details), while the black solid lines correspond to the fitted results of the Hagan-SABR model.

As shown in Fig.~\ref{skew-fig16}, the SABR model exhibits systematic fitting discrepancies across different underlying assets and maturities. In particular, in the right wing corresponding to OTM call options, the model-implied volatilities deviate significantly from the market values, with little overlap between the two curves. The fitting performance is relatively satisfactory around the at-the-money region and in the left wing corresponding to OTM put options. These results indicate that the standard Hagan-SABR model has limitations in capturing the asymmetric structure of the implied volatility surface in the domestic index options market. This limitation motivates the need for model extensions beyond the standard Hagan-SABR framework to better capture the observed asymmetry in implied volatility curves.

\section{An Improved Skew-Enhanced SABR Model}

\label{skew-sec3}

\subsection{Model Specification and Derivation of Implied Volatility Formula}

As indicated by the empirical results presented earlier, the standard SABR model exhibits relatively low fitting accuracy when applied to implied volatility curves. Therefore, it is necessary to introduce appropriate modifications to the SABR framework, while preserving its structural simplicity, in order to enhance its ability to capture observed market implied volatility patterns. 
To this end, this paper proposes an improved approach that significantly enhances the fitting performance by introducing only one or two  additional degrees of freedom , while maintaining a  parsimonious analytical form of the resulting implied volatility approximation, thereby ensuring computational efficiency in practical applications.

Specifically, within the original two-factor SABR framework, instead of directly modeling the volatility process $\hat{\alpha}$, we model the dynamics of its square $\hat{\alpha}^2$, i.e., the variance process. That is, we assume
\begin{equation}
d\hat{\alpha}^2 = \nu \hat{\alpha}^2 dW_2 ,
\label{14}
\end{equation}
where $\nu$ represents the volatility of the variance factor.

However, directly adopting the above specification would render the associated stochastic differential equations analytically intractable for deriving a closed-form or asymptotic implied volatility expression. To retain the applicability of the asymptotic expansion framework developed by Hagan et al., we introduce an appropriate change of variables such that the resulting stochastic differential equations can still be expressed in a form consistent with Eq.~(\ref{skew-2}). 
This transformation allows the derivation of an analytical approximation for the implied volatility under the modified model.

Let $\theta = \hat{\alpha}^2$, then $d \theta = \nu \theta dW_2$. Define $Y_t = Y_t(\theta, t) = \theta^{\frac{1}{2}}e^{\frac{1}{8}\nu^2t}$. Computing the partial derivatives yields,
\begin{align*}
  \frac{\partial Y_t}{\partial t} &= \frac{1}{8}\nu^2 \theta^{\frac{1}{2}}e^{\frac{1}{8}\nu^2t} = \frac{1}{8}\nu^2 Y_t  \\ \frac{\partial Y_t}{\partial \theta} &= \frac{1}{2}\theta^{-\frac{1}{2}}e^{\frac{1}{8}\nu^2t} = \frac{1}{2}\theta^{-1}Y_t \\ \frac{\partial^2 Y_t}{\partial \theta^2} &= -\frac{1}{4}\theta^{-\frac{3}{2}}e^{\frac{1}{8}\nu^2t} = -\frac{1}{4}\theta^{-2}Y_t
\end{align*}

By It\^o's lemma,
\begin{align*}
    dY_t &= \frac{\partial Y_t}{\partial t}dt + \frac{\partial Y_t}{\partial \theta}d\theta + \frac{1}{2}\frac{\partial^2 Y_t}{\partial \theta^2}d\theta^2 \\ &= \frac{1}{8}\nu^2 Y_t dt + \frac{1}{2}\theta^{-1}Y_t \nu \theta dW_2 + \frac{1}{2}(-\frac{1}{4}\theta^{-2}Y_t) \nu^2 \theta^2 dt \\ &= \frac{1}{2}\nu Y_t dW_2
\end{align*}

In this way, we obtain a stochastic process $Y_t$ that is of the same order of magnitude as the volatility-like coefficient $\hat{\alpha}_t$. However, $Y_t$ cannot be directly interpreted as the volatility process in the model, and therefore an appropriate rescaling is required. To this end, a level adjustment term is introduced based on $Y_t$, denoted by $\gamma \cdot Y_t$, where $\gamma = \frac{\alpha + m}{\alpha}$, which serves as a coefficient to adjust the level of $Y_t$. As a result, we have $d\hat{F} = \gamma \cdot Y_t \hat{F}dW_1$. Hence, the final model can be written as
\begin{equation}
\left\{
\begin{aligned}
& \quad d\hat{F} = Y_t C(F)dW_1, \quad \hat{F}(0) = f, \\
&\quad dY_t = \frac{1}{2}\nu Y_tdW_2, \quad Y_0 = \hat{\alpha}(0) = \alpha, \\ &\quad dW_1dW_2 = \rho dt.   \label{skew-19}
\end{aligned}
\right.
\end{equation}
where $C(F) = \frac{\alpha + m}{\alpha}\hat{F}$. This model shares the same functional form as Eq.~(\ref{skew-2}), and therefore the solution formulas provided by Hagan, namely Eq.~(\ref{skew-3}) and Eq.~(\ref{skew-4a})$\sim$(\ref{skew-4d}), can be directly applied to compute $I_0$ and $I_1$. Let $u = \frac{1}{2}\nu$. We first show that the solution to the above system can be approximated by $\sigma_B^{H\text{-}SABR}(\alpha+m, \rho, u; K, \tau)$, whose explicit expression is given in Eq.~\eqref{skew-12}.

For notational simplicity, in the following, whenever no ambiguity arises, the dependence of the Hagan-SABR solution $\sigma_B^{H\text{-}SABR}$ on $K$ and $\tau$ will be omitted, and it will be denoted in the form $\sigma_B^{H\text{-}SABR}(\alpha, \rho, \nu)$.

Substituting $C(\hat{F}) = \gamma \hat{F}^{\beta}$ into Eq.~(\ref{skew-4a})$\sim$(\ref{skew-4d}) and setting $\beta = 1$. Since
\begin{equation*}
    C(F) = \gamma F^{\beta}, \quad C'(F) = \gamma \beta F^{\beta-1}, \quad C''(F) = \gamma \beta (\beta-1)F^{\beta-2}.
\end{equation*}
it follows that
\begin{align*}
 \gamma_1 = \frac{C'(f_{ave})}{C(f_{ave})} &= \frac{\gamma \beta f_{ave}^{\beta-1}}{\gamma f_{ave}^{\beta}} = \beta f_{ave}^{-1}, \\ \gamma_2 = \frac{C''(f_{ave})}{C(f_{ave})} &= \frac{\gamma \beta (\beta -1)f_{ave}^{\beta-2}}{\gamma f_{ave}^{\beta}} = \beta (\beta - 1)f_{ave}^{-2} \\ 2\gamma_2-\gamma_1^2 + 1/f_{ave}^2 &= (\beta - 1)^2 f_{ave}^{-2}.
\end{align*}

When $\beta = 1$, the above term vanishes. This is consistent with the standard Hagan-SABR model, where the corresponding term is also equal to zero when $\beta = 1$. For the remaining part of $I_1$, we have
\begin{equation*}
    \frac{1}{4}\rho u \alpha \gamma_1 C(f_{ave}) = \frac{1}{4}\rho u \alpha \gamma \beta f_{ave}^{\beta-1} = \frac{1}{4}\rho u \gamma \alpha
\end{equation*}
Substituting $\gamma = \frac{\alpha + m}{\alpha}$ into the above expression yields
\begin{equation*}
    \frac{1}{4}\rho u \alpha \gamma_1 C(f_{ave}) = \frac{1}{4}\rho u (\alpha + m)
\end{equation*}
Therefore, the $I_1$ term in Eq.~(\ref{skew-19}) can be expressed as\begin{equation*}
    I_1 = \frac{1}{4}\rho u (\alpha + m) + \frac{2-3\rho^2}{24}u^2 = I_1^H(\alpha+m, \rho, u)
\end{equation*}
This expression is structurally identical to the $I_1$ term in the Hagan-SABR model, with the only difference being that the parameter $\alpha$ is replaced by $\alpha + m$ in the first argument.

For the $I_0$ term, we have
\begin{align*}
    &\int_K^f \frac{dx}{C(x)} = \int_K^f \frac{dx}{\gamma x} = \frac{1}{\gamma}\log \frac{f}{K} \\ &\zeta = \frac{u}{\alpha}\cdot \frac{\sqrt{fK} \log \frac{f}{K}}{\gamma \sqrt{fK}} = \frac{1}{\gamma} \frac{u}{\alpha} \log \frac{f}{K}
\end{align*}

Thus, it follows that
\begin{align*}
    I_0 &= \frac{\alpha \log \frac{f}{K}}{\int_K^f \frac{dx}{C(x)}} \cdot  \bigg (\frac{\zeta}{\chi(\zeta)}\bigg ) = \frac{\alpha \log \frac{f}{K}}{\frac{1}{\gamma}\log \frac{f}{K}} \cdot   \frac{\frac{1}{\gamma} \frac{u}{\alpha} \log \frac{f}{K}}{\chi(\zeta)} = \frac{u \log \frac{f}{K}}{\chi(\frac{u \log \frac{f}{K}}{\gamma \alpha})}    \\ &= \frac{u \log \frac{f}{K}}{\chi(\frac{u \log \frac{f}{K}}{\alpha+m})} = I_0^H(\alpha+m, \rho, u)
\end{align*}

Therefore, the $I_0$ and $I_1$ terms of the equation coincide exactly with $I_0^H(\alpha + m, \rho, u)$ and $I_1^H(\alpha + m, \rho, u)$, respectively. Hence, the solution to Eq.~(\ref{skew-19}) can be written as
\[
I_0 \cdot (1 + I_1 \tau) = I_0^H(\alpha + m, \rho, u) \cdot (1 + I_1^H(\alpha + m, \rho, u) \tau) = \sigma_B^{H\text{-}SABR}(\alpha + m, \rho, u).
\]

At this point, we have proposed the new model in Eq.~(\ref{skew-19}) and shown that its solution can be expressed as $\sigma_B = \sigma_B^{H-SABR}(\alpha+m, \rho, u) + error_{sol}$, where $error_{sol}$ arises from the approximation error in solving the model. More specifically, this error originates from the asymptotic approximation formula provided by Hagan et al. The model thus involves four unknown parameters: $\alpha, \rho, \nu, m$.

To further analyze these parameters and obtain a more concise analytical representation, we proceed to expand the expression $\sigma_B^{H-SABR}(\alpha, \rho, \nu)$. First, we perform a Taylor expansion of $\chi(z)$:
\[
\chi(z) = \chi(0) + \chi'(0) z + \frac{1}{2} \chi''(0) z^2 + \frac{1}{6} \chi^{(3)}(0) z^3 + {\scriptscriptstyle\mathcal{O}}(z^4).
\]
where
\begin{equation*}
  \chi(0) = 0, \chi'(0) = 1, \chi''(0) = \rho , \frac{1}{6}\chi^{(3)}(0) = 3\rho^2 - 1
\end{equation*}

Accordingly, by performing a Taylor expansion of $I_0^{H}(\alpha, \rho, \nu)$, we obtain a second-order approximation with the associated truncation error given by
\begin{align*}
    I_0^{H}(\alpha, \rho, \nu) &= \frac{\alpha \zeta}{\chi(\zeta)} = \frac{\alpha \zeta}{\zeta + \frac{1}{2}\rho \zeta^2 + \frac{1}{6}(3\rho^2 - 1)\zeta^3+{\scriptscriptstyle\mathcal{O}}(\zeta^3)} \\ &= \frac{\alpha }{1 + \frac{1}{2}\rho \zeta + \frac{1}{6}(3\rho^2 - 1)\zeta^2+{\scriptscriptstyle\mathcal{O}}(\zeta^2)} \\ &= \alpha \big(1 - \frac{1}{2}\rho \zeta - \frac{1}{6}(3\rho^2 - 1)\zeta^2 + \frac{1}{4}\rho^2\zeta^2\big) +{\scriptscriptstyle\mathcal{O}}(\zeta^2) \\ &= \alpha \big(1 - \frac{1}{2}\rho \zeta +(\frac{1}{6} - \frac{1}{4}\rho^2)\zeta^2\big) +{\scriptscriptstyle\mathcal{O}}(\zeta^2)
\end{align*}

Substituting $\zeta^H(\alpha, \rho, \nu) = \frac{\nu}{\alpha} \log \frac{f}{K} = -\frac{\nu}{\alpha} \log \frac{K}{f}$, we have\begin{equation}
    I_0^{H}(\alpha, \rho, \nu) = \alpha + \frac{1}{2}\rho \nu \log \frac{K}{f} + (\frac{1}{6}-\frac{1}{4}\rho^2)\frac{\nu^2}{\alpha}\log^2\frac{K}{f} + {\scriptscriptstyle\mathcal{O}}(\log^2\frac{K}{f}) \label{11}
\end{equation}

Therefore, the SABR formula derived by Hagan can be written as\begin{align*}
   &\sigma_B^{H-SABR}(\alpha, \rho, \nu)= I_0^H(\alpha, \rho, \nu) \cdot \big(1 + I_1^H(\alpha, \rho, \nu) \tau\big) \\ &= \frac{\alpha \zeta}{\chi(\zeta)} \cdot \big(1 + (\frac{1}{4}\rho \nu \alpha + \frac{2-3\rho^2}{24}\nu^2)\tau \big) \\ &= \big[\alpha + \frac{1}{2}\rho \nu \log \frac{K}{f} + (\frac{1}{6}-\frac{1}{4}\rho^2)\frac{\nu^2}{\alpha}\log^2\frac{K}{f} + {\scriptscriptstyle\mathcal{O}}(\log^2\frac{K}{f})\big]\cdot \big(1 + (\frac{1}{4}\rho \nu \alpha + \frac{2-3\rho^2}{24}\nu^2)\tau \big)
\end{align*}

It can be observed from the above expressions that adjusting $\alpha$ affects the constant term of the curve, while modifying $\nu$ changes the coefficient of the linear term, i.e., the skew component. To more clearly characterize the impact of each parameter on the functional form, we consider applying a shift transformation to the parameters. For any constant $m$, we have
\begin{align*}
   I_0^H(\alpha+m, \rho, \nu) &= I_0^H(\alpha, \rho, \nu) + m + {\mathcal{O}}(\log^2\frac{K}{f}) \\  I_1^H(\alpha+m, \rho, \nu) &= I_1^H(\alpha, \rho, \nu) + \frac{1}{4}\rho\nu m
\end{align*}
Thus, \begin{equation}
\begin{aligned}
   \sigma_B^{H-SABR}(\alpha+m, \rho, \nu)  &= I_0^H(\alpha+m, \rho, \nu) \cdot \big[1 + I_1^H(\alpha+m, \rho, \nu)\tau\big] \\ &= \big[I_0^H(\alpha, \rho, \nu) + m + {\mathcal{O}}(\log^2\frac{K}{f}) \big] \cdot \big[1 + (I_1^H(\alpha, \rho, \nu) + \frac{1}{4}\rho\nu m)\tau  \big] \\ &= \sigma_B^{H-SABR}(\alpha, \rho, \nu) + m + m\cdot I_1^H(\alpha+m, \rho, \nu)\tau\\  &+ \frac{1}{4}\rho\nu m (\alpha +\frac{1}{2}\rho \nu \log\frac{K}{f})\tau  + {\mathcal{O}}(\log^2\frac{K}{f})
\end{aligned} \label{skew-17}
\end{equation}

Since $m\cdot[I_1^H(\alpha+m, \rho, \nu) + \frac{1}{4}\rho \nu(\alpha +\frac{1}{2}\rho \nu \log\frac{K}{f})]\tau\ll1$, we have the approximation
\begin{equation*}
    \sigma_B^{H-SABR}(\alpha+m, \rho, \nu; K, \tau) \approx \sigma_B^{H-SABR}(\alpha, \rho, \nu; K, \tau) + m+ {\mathcal{O}}(\log^2\frac{K}{f})
\end{equation*}
which implies that adjusting $\alpha$ effectively shifts the overall level of the implied volatility curve.

For any constant $w$, we have
\begin{equation*}
   I_0^H(\alpha, \rho, \nu+w) = I_0^H(\alpha, \rho, \nu) + \frac{1}{2}\rho w \log \frac{K}{f} + {\mathcal{O}}(\log^2\frac{K}{f})
\end{equation*}
\begin{equation*}
    I_1^H(\alpha, \rho, \nu+w) = I_1^H(\alpha, \rho, \nu) +  \frac{1}{4}\rho w \alpha + \frac{2-3\rho^2}{24}(2\nu w+w^2)
\end{equation*}

Let\begin{equation*}
    \Delta I_1^H(\alpha, \rho, \nu; w) = I_1^H(\alpha, \rho, \nu+w) - I_1^H(\alpha, \rho, \nu) = \frac{1}{4}\rho w \alpha + \frac{2-3\rho^2}{24}(2\nu w+w^2)
\end{equation*}

Therefore\begin{equation}
\begin{aligned}
 \sigma_B^{H-SABR}(\alpha, \rho, \nu+w)  &=  I_0^H(\alpha, \rho, \nu+w) \cdot \big[1 + I_1^H(\alpha, \rho, \nu+w)\tau\big] \\ &= \big[I_0^H(\alpha, \rho, \nu) + \frac{1}{2}\rho w \log \frac{K}{f} + {\mathcal{O}}(\log^2\frac{K}{f}) \big] \cdot \big[1 + I_1^H(\alpha, \rho, \nu+w)\tau  \big]\\ & =\sigma_B^{H-SABR}(\alpha, \rho, \nu) + I_0^H(\alpha, \rho, \nu)\cdot \Delta I_1^H(\alpha, \rho, \nu; w) \cdot \tau\\  &+ \frac{1}{2}\rho w \log \frac{K}{f}  + \frac{1}{2}\rho w\log \frac{K}{f} \cdot I_1^H(\alpha, \rho, \nu+w)\cdot \tau + {\mathcal{O}}(\log^2\frac{K}{f})
\end{aligned} \label{skew-18}
\end{equation}

Since $\Delta I_1^H(\alpha, \rho, \nu; w) \cdot \tau \ll 1, I_1^H(\alpha, \rho, \nu+w)\cdot \tau \ll 1$, we have the approximation
\begin{equation*}
    \sigma_B^{H-SABR}(\alpha, \rho, \nu+w; K, \tau) \approx \sigma_B^{H-SABR}(\alpha, \rho, \nu; K, \tau) + \frac{1}{2}\rho w \log \frac{K}{f}+ {\mathcal{O}}(\log^2\frac{K}{f})
\end{equation*}
Therefore, adjusting $\nu$ introduces an explicit skew component into the implied volatility formula.

Substituting Eq.~(\ref{skew-17}) and Eq.~(\ref{skew-18}), the implied volatility of Eq.~(\ref{skew-19}) can be approximated as\begin{align*}
    \sigma_B &= \sigma_B^{H-SABR}(\alpha+m, \rho, u) + error_{sol} \\ &= \sigma_B^{H-SABR}(\alpha, \rho, u)+m + m\cdot I_1^H(\alpha+m, \rho, u)\tau\\  &+ \frac{1}{4}\rho u m (\alpha+\frac{1}{2}\rho u \log\frac{K}{f})\tau  + {\mathcal{O}}(\log^2\frac{K}{f})+error_{sol}  \\ &= \sigma_B^{H-SABR}(\alpha, \rho, \nu)+ I_0^H(\alpha, \rho, \nu)\cdot \Delta I_1^H(\alpha, \rho, \nu; w) \cdot \tau \\ &+ \frac{1}{2}\rho w\log \frac{K}{f} + \frac{1}{2}\rho w\log \frac{K}{f} \cdot I_1^H(\alpha, \rho, u)\tau+m+ m\cdot I_1^H(\alpha+m, \rho, u)\tau \\&+ \frac{1}{4}\rho u m (\alpha+\frac{1}{2}\rho u \log\frac{K}{f})\tau  + {\mathcal{O}}(\log^2\frac{K}{f})+error_{sol}
\end{align*}

Let\begin{align}
    c^* &= \frac{1}{2}\rho w +  +\big[\frac{1}{2}\rho wI_1^H(\alpha, \rho, u) +\frac{1}{8}\rho^2 u^2m + \frac{1}{2}\rho \nu \Delta I_1^H(\alpha, \rho, \nu; w) \big]\tau\label{skew-20} \\ d^* &= m + \big[m\cdot I_1^H(\alpha+m, \rho, u)+\frac{1}{4}\alpha \rho u m+\alpha \cdot \Delta I_1^H(\alpha, \rho, \nu; w) \big] \tau  \label{skew-21}
\end{align}

Thus\begin{equation*}
    \sigma_B = \sigma_B^{H-SABR}(\alpha, \rho, \nu) + c^*\cdot \log \frac{K}{f} + d^* + {\mathcal{O}}(\log^2\frac{K}{f})+error_{sol}
\end{equation*}

The observed market implied volatility can be written as
\begin{equation*}
    \sigma_{market} = \sigma_B + error_{model} = \sigma_B^{H-SABR}(\alpha, \rho, \nu) + c^*\cdot \log \frac{K}{f} + d^* + {\mathcal{O}}(\log^2\frac{K}{f})+error_{all}
\end{equation*}
where $error_{model}$ denotes the error introduced by the model specification in Eq.~(\ref{skew-19}), for example, the approximation error arising from approximating the true dynamics of $\hat{F}$ by $d\hat{F} = Y_tC(F)dW_1$. Therefore, the total error can be expressed as
\begin{equation*}
error_{all} = error_{model} + error_{sol}.
\end{equation*}
that is, the overall error consists of both model specification error and solution (approximation) error.

Thus far, only one additional degree of freedom, $m$, has been introduced into the model. However, since the expressions for $c^*$ and $d^*$ are relatively complicated, they are not convenient for direct use in practical calibration. To simplify the model representation and improve the tractability of parameter calibration, we further reparameterize the model by introducing two new degrees of freedom, $c$ and $d$, to replace $c^*$ and $d^*$, which are implicitly determined by $m$. The model can thus be rewritten as
\begin{equation*}
    \sigma_B = \sigma_B^{H-SABR}(\alpha, \rho, \nu) + c\cdot \log \frac{K}{f} + d
\end{equation*}

Under this specification, $c$ and $d$ are treated as new model parameters to be calibrated. Compared with the original formulation, this reparameterization offers two main advantages. First, by explicitly introducing the coefficients of the constant and linear terms, the model is able to more flexibly capture the skewness structure of the implied volatility curve. Second, since the model already contains free parameters associated with the constant and linear terms in $\log\frac{K}{f}$, the corresponding low-order deviations can be absorbed by $c$ and $d$ during calibration. In other words, by explicitly modeling the constant and linear components, the model can automatically match the lower-order terms in the implied volatility expansion, thereby shifting the dominant approximation error to higher-order terms. Consequently, the remaining approximation error is primarily determined by higher-order terms and can be controlled at the order of ${\mathcal{O}}\!\left(\log^2\frac{K}{f}\right)$.

Based on the above improvement, since the resulting expression explicitly contains coefficients that characterize skewness, we refer to this model as the skew-SABR model. In fact, as indicated by Eq.~(\ref{skew-18}), it is precisely the shift from modeling volatility to modeling variance that leads to a structural change in the skew term, thereby enabling the final expression to explicitly capture skewness. Therefore, the name skew-SABR accurately reflects the core modeling idea, namely, extending from volatility modeling to variance-based modeling.

Under the above model specification, the corresponding implied volatility approximation is denoted by $\sigma_B^{skew}$, i.e.,
\begin{align}
   \sigma_B^{skew} &= \sigma_B^{H-SABR}(\alpha, \rho, \nu) + c\cdot \log \frac{K}{f} + d \label{skew-23} \\  \sigma_{market} = \sigma_B^{skew} +  {\mathcal{O}}(\log^2\frac{K}{f}) &= \sigma_B^{H-SABR}(\alpha, \rho, \nu) + c\cdot \log \frac{K}{f} + d + {\mathcal{O}}(\log^2\frac{K}{f}) \notag
\end{align}

At this stage, the construction of the skew-SABR model is complete. The model involves five parameters to be calibrated: $\alpha$, $\rho$, $\nu$, $c$, and $d$. Compared with the original Hagan-SABR model, which requires four parameters, the skew-SABR model introduces only one additional degree of freedom. As demonstrated by the empirical results in Sec.~\ref{skew-sec4}, this parsimonious modification leads to a significant improvement in fitting accuracy for market implied volatility surfaces.

\subsection{Consistency between Model Specification and Market Data}

It should be emphasized that, although the implied volatility formula of the skew-SABR model in Eq.~(\ref{skew-23}) appears to differ from the Hagan-SABR implied volatility formula $\sigma_B^{H\text{-}SABR}$ only by the addition of two terms, $c \cdot \log \frac{K}{f} + d$, its essence is not merely a correction to the original model error $\sigma_{market} - \sigma_B^{H\text{-}SABR}$. Instead, it represents a reformulation of the underlying stochastic model structure. The corresponding stochastic process is given by Eq.~(\ref{skew-19}), rather than Eq.~(\ref{skew-10}), which underlies the Hagan-SABR formula.
Therefore, in Eq.~(\ref{skew-23}), although the first term $\sigma_B^{H\text{-}SABR}$ retains the same functional form as in the Hagan-SABR model, the interpretation of its parameters has fundamentally changed. For instance, in the skew-SABR model, the parameter $\nu$ characterizes the volatility of the variance process, while $\alpha + d$ represents the overall volatility level. These interpretations differ significantly from the financial meanings of the corresponding parameters in the original SABR framework.

This subsection examines the consistency between the model specification and market data. First, a numerical experiment is conducted to compare the theoretically derived parameter $c^*$ with the optimal parameter $c$ obtained by calibrating the implied volatility formula in Eq.~(\ref{skew-23}) to market data. A close agreement between the two would, on the one hand, support the validity of the stochastic process specified in Eq.~(\ref{skew-19}), and on the other hand, verify the accuracy of the derivation of the implied volatility approximation in Eq.~(\ref{skew-23}). This provides empirical support for both the model specification and the correctness of its approximate solution.

Furthermore, the calibrated model parameters are analyzed in terms of their numerical characteristics to assess whether they are consistent with their intended financial interpretations. For example, we examine whether the parameter $\nu$ effectively captures the volatility of the variance process, and whether $\alpha + d$ appropriately represents the volatility level.
Through these two aspects of analysis, this subsection provides a systematic evaluation of the consistency between the model specification and market data from both the perspectives of model structure and economic interpretation.

First, we examine the consistency between the theoretical value $c^*$ and the calibrated parameter $c$ through numerical experiments. According to Eq.~(\ref{skew-20}), when 
\[
\bigg[\frac{1}{2}\rho w I_1^H(\alpha, \rho, \nu) + \frac{1}{8}\rho^2 \nu^2 m + \frac{1}{2}\rho \nu \Delta I_1^H(\alpha, \rho, \nu; w) \bigg]\tau \ll 1,
\]
an approximate relationship can be obtained:
\[
c^* \approx \frac{1}{2}\rho w = -\frac{1}{4}\rho \nu.
\]
Accordingly, $-\frac{1}{4}\rho \nu$ is taken as the theoretical value of the parameter $c$ for comparison.

Taking the June 2025 contract of the E Fund ChiNext ETF options (denoted as XSHE\_159915:202506) as an example, we use the full sample of market data over its listing period, from October 24, 2024 to June 24, 2025. Model parameters are calibrated at 5-minute intervals. The results show that, over this period, the mean of the calibrated parameter $c$ is $-0.040$, while the mean of the theoretical value $c^*$ is $-0.035$. The correlation coefficient between the two reaches $0.964$, indicating a high degree of consistency in both level and dynamic variation.

\begin{figure}[htbp]
    \centering
    \includegraphics[width=0.8\textwidth]{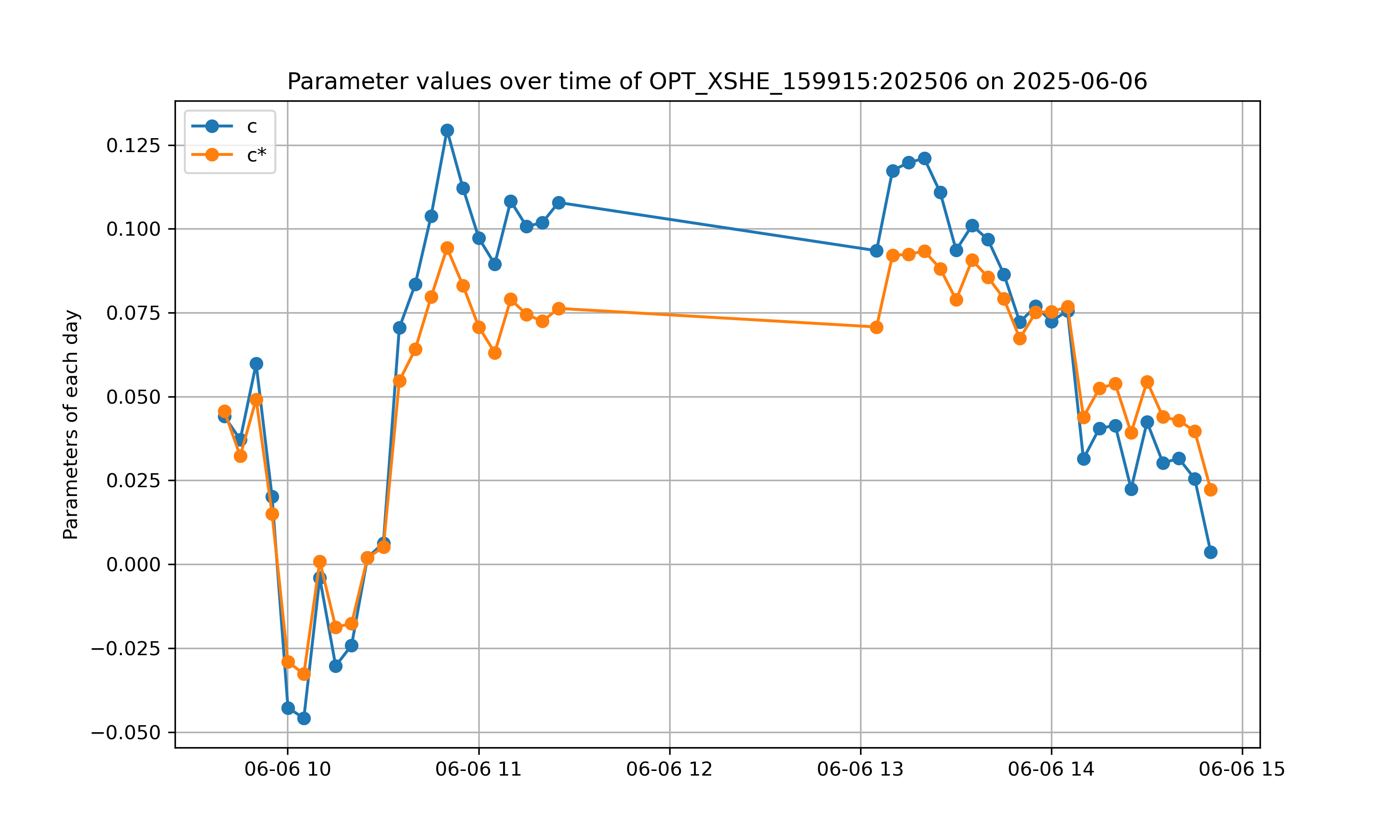}
  \caption{Time series comparison of the calibrated parameter $c$ and the theoretical value $c^*$ for the XSHE\_159915:202506 contract on June 6, 2025} \label{skew-fig11}
\end{figure}

Furthermore, a representative trading day (June 6, 2025) is selected, and the time series of $c$ and $c^*$ are plotted, as shown in Fig.~\ref{skew-fig11}.

From Fig.~\ref{skew-fig11}, it can be observed that $c$ and $c^*$ exhibit a high degree of consistency in their dynamic behavior. Their time series evolve almost synchronously, with very close numerical values. This result indicates that the parameter $c$ obtained via calibration of the implied volatility formula is highly consistent with the theoretically derived value $c^*$. This empirical finding not only supports the validity of the proposed stochastic process, but also verifies the effectiveness of the implied volatility approximation derived from it.

On the other hand, we compare the parameters calibrated under the skew-SABR model with their counterparts in the Hagan-SABR model. Seven option products are considered, including CSI 1000 equity index (MO) options, CSI 300 equity index (IO) options, SSE 50 equity index (HO) options, CSI 300 ETF (510300) options, CSI 500 ETF (510500) options, SSE 50 ETF (510050) options, and E Fund ChiNext ETF (159915) options, all with the June 2025 maturity. The full listing period of each contract is examined.

For index options (MO, IO, and HO), the June 2025 contracts are listed from June 24, 2024, to June 20, 2025, covering approximately one year. For ETF options, the corresponding contracts are listed from October 24, 2024, to June 24, 2025, spanning about eight months. As the June contracts are quarterly contracts, using their full listing period captures a relatively long market horizon, thereby enhancing the representativeness of the empirical results.

For each trading day, data within the first and last 10 minutes of trading are excluded. Observations are sampled every 5 minutes between 9:40:05 and 14:50:05, and both the skew-SABR and Hagan-SABR models are calibrated separately. Consistent with earlier findings, the Hagan-SABR model fails to accurately fit the market implied volatility curve even under optimal parameters, whereas the skew-SABR model achieves a significantly better fit. A detailed comparison of the fitting performance will be provided in Section~\ref{skew-sec4}; here, we focus on differences in the calibrated parameters.

Several representative findings emerge from the numerical experiments. First, the parameter $\nu$ estimated under the skew-SABR model is consistently and significantly larger than its counterpart in the Hagan-SABR model. Specifically, for each trading day, the ratio $\nu_{\text{skew}} / \nu_{\text{Hagan}}$ is computed at each time point, and its intraday average is taken as the daily value.
During the sample period, the June 2025 contracts experienced episodes of heightened market volatility (e.g., major market rallies and external shocks), which may lead to extreme estimates of $\nu$. Therefore, observations outside the $5\%$ and $95\%$ quantiles are excluded, and the mean of the remaining samples is reported.

Table~\ref{skew-table1} summarizes the ratios of $\nu$ estimated by the two models across different products. The results show that the $\nu$ parameter in the skew-SABR model is approximately $1.6$ to $2.6$ times that of the Hagan-SABR model, with an average of about $2$. This indicates that the interpretation of $\nu$ in the skew-SABR model has changed: it no longer corresponds to the volatility in the original SABR setting, but instead characterizes the volatility of the variance process. This is consistent with the economic interpretation of $\nu$ under the proposed model specification.

\begin{table}[H]
    \centering
    \caption{Ratio of the parameter $\nu$ estimated by the skew-SABR and Hagan-SABR models across different option products}
    \begin{tabular}{|c|c|c|c|c|c|c|c|c|}
    \hline
        symbol & MO & IO & HO & 510300 & 510500 & 510050 & 159915 & Average  \\ \hline
        $\nu_{skew}$/$\nu_{Hagan}$ & 2.69  & 1.87  & 1.91  & 2.00  & 1.31  & 2.07  & 1.62  & 1.92  \\ \hline
    \end{tabular}
    \label{skew-table1}
\end{table}

In addition, in the Hagan-SABR model, the quantity $\alpha \cdot (1 + I_1 \cdot \tau)$ characterizes the level of the implied volatility curve in the vicinity of at-the-money options. In the skew-SABR model, this level is given by $\alpha \cdot (1 + I_1 \cdot \tau) + d$, i.e., an additional level adjustment term is introduced to the original structure. Numerical results show that the quantity $\alpha \cdot (1 + I_1 \cdot \tau) + d$ obtained from the skew-SABR model is numerically very close to $\alpha \cdot (1 + I_1 \cdot \tau)$ from the Hagan-SABR model.

\begin{figure}[htbp]
    \centering
\includegraphics[width=0.9\textwidth]{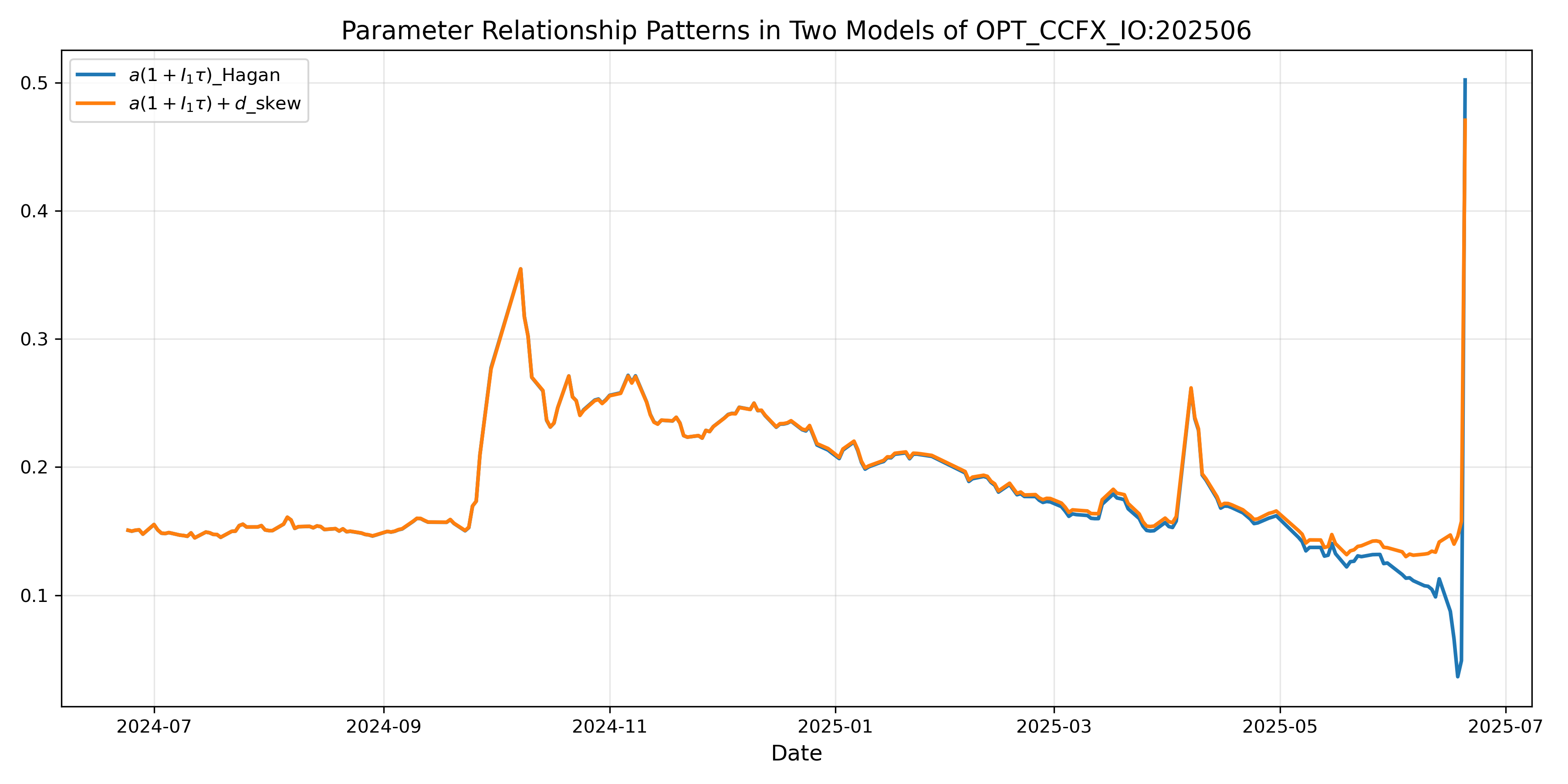}\\
\caption{Comparison of $\alpha \cdot (1 + I_1\cdot \tau) + d$ from the skew-SABR model and $\alpha \cdot (1 + I_1\cdot \tau)$ from the Hagan-SABR model for CSI 300 equity index options} \label{skew-fig24}
\end{figure}

\begin{figure}[htbp]
    \centering
\includegraphics[width=0.9\textwidth]{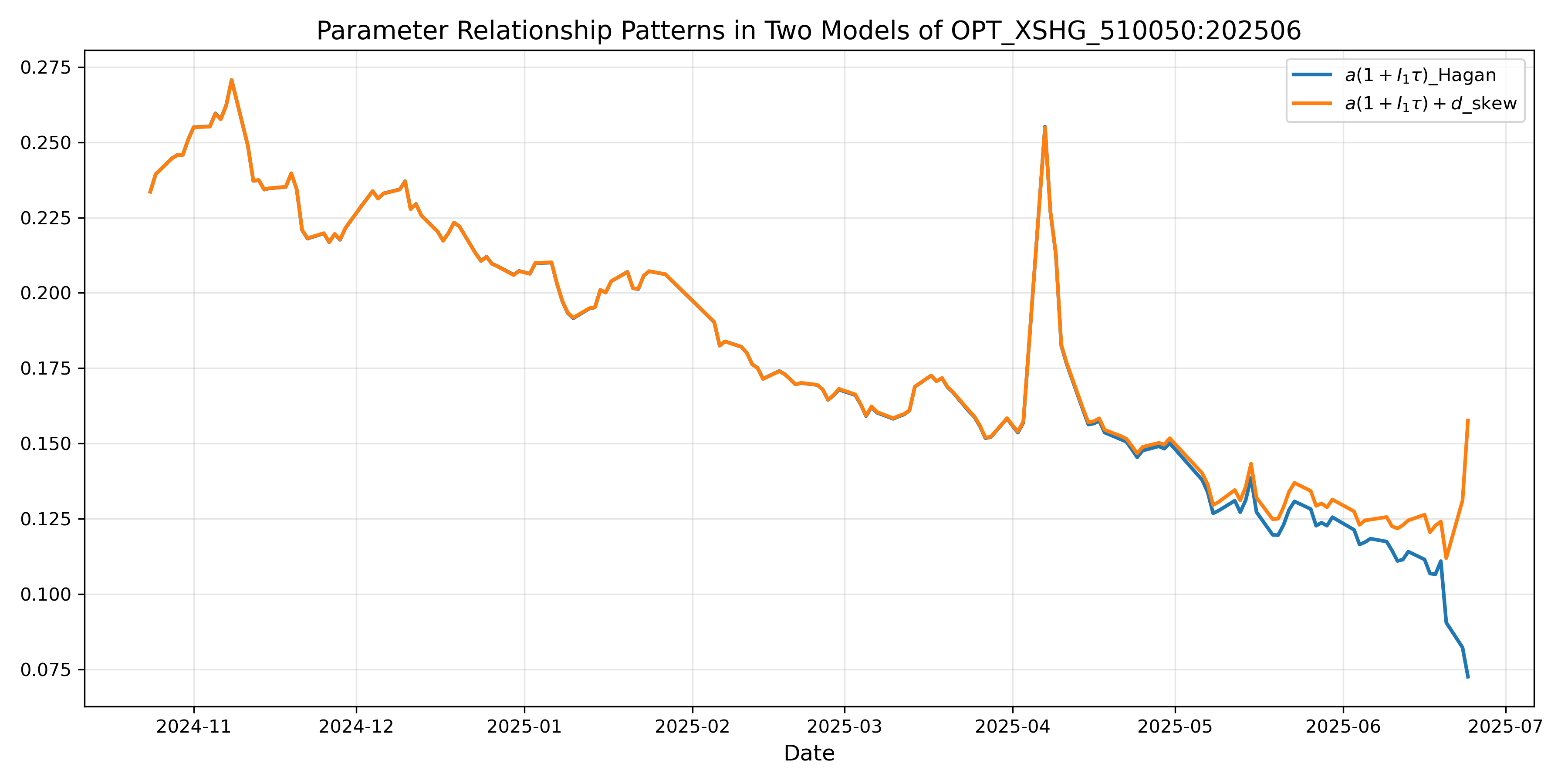}\\
\caption{Comparison of $\alpha \cdot (1 + I_1\cdot \tau) + d$ from the skew-SABR model and $\alpha \cdot (1 + I_1\cdot \tau)$ from the Hagan-SABR model for SSE 50 ETF options} \label{skew-fig25}
\end{figure}

Figs.~\ref{skew-fig24} and~\ref{skew-fig25} illustrate the time series of these two quantities for CSI 300 index options and SSE 50 ETF options, respectively. The two series exhibit a high degree of consistency over time. This empirical finding not only validates the introduction of the level adjustment term $m$ in the stochastic process Eq.~(\ref{skew-19}) for the parameter $Y_t$, but also provides market-based support for the theoretical relation in Eq.~(\ref{skew-21}), namely that $d^*$ is approximately equal to the level adjustment term $m$. Therefore, using the parameter $d$ rather than directly using $m$ in the final implied volatility expression preserves the parsimony of $\sigma_{skew}$ while fully capturing the structural adjustment induced by the level correction.

In summary, although both the Hagan-SABR and skew-SABR models include the parameters $\alpha$ and $\nu$, the calibrated values of these identically named parameters differ significantly across the two models, indicating that their financial interpretations are not consistent. Therefore, the skew-SABR model should not be viewed merely as an error correction to the Hagan-SABR implied volatility formula. The additional term $c \cdot \log \frac{K}{f} + d$ is not introduced to capture the fitting error $\sigma_{market} - \sigma_B^{H\text{-}SABR}$, but fundamentally reflects a modification of the underlying stochastic process, namely the model proposed in Eq.~(\ref{skew-19}).

Furthermore, in the implied volatility representation, the parameter $m$ introduced in model~(\ref{skew-19}) is extended into two parameters, $c$ and $d$. This formulation remains consistent with the analytical solution of the stochastic process and, by separately capturing the level and skew components of the volatility curve, enables the first-order structure originally embedded in the error term $error_{all}$ to be explicitly incorporated into the implied volatility formula. As a result, the systematic deviation between model-implied and market-implied volatility is significantly reduced.

Based on the above analysis, the numerical experiments in this section validate the rationality of the model specification and the effectiveness of the implied volatility approximation from both parameter consistency and economic interpretation perspectives. The next subsection will further illustrate, through graphical analysis, the impact of each parameter on the shape of the implied volatility curve.

\subsection{Impact of Model Parameters on the Implied Volatility Curve}

We restate the implied volatility formula of the skew-SABR model, denoted by $\sigma^{\text{skew}}_B$, and perform a Taylor expansion of the term $\frac{\zeta}{\chi(\zeta)}$:
\begin{equation*}
\begin{aligned}
 \sigma_B^{skew}&(\alpha, \rho, \nu, c, d; K, \tau) = \sigma_B^{H-SABR}(\alpha, \rho, \nu) + c\cdot \log \frac{K}{f} + d \\ &= \frac{\alpha \zeta}{\chi(\zeta)} \cdot \big(1 + (\frac{1}{4}\rho \nu \alpha + \frac{2-3\rho^2}{24}\nu^2)\tau \big) + c\cdot \log \frac{K}{f} + d \\ &= \big[\alpha + \frac{1}{2}\rho \nu \log \frac{K}{f} + (\frac{1}{6}-\frac{1}{4}\rho^2)\frac{\nu^2}{\alpha}\log^2\frac{K}{f} + {\scriptscriptstyle\mathcal{O}}(\log^2\frac{K}{f})\big]\cdot \big(1 + I_1^H(\alpha, \rho, \nu)\tau \big) + c\cdot \log \frac{K}{f} + d
\end{aligned}
\end{equation*} 

The above expansion reveals that different parameters govern distinct geometric features of the implied volatility curve. Specifically, $\alpha + d$ primarily determines the overall level of the volatility curve, corresponding to the volatility level around at-the-money options. The term $\frac{1}{2}\rho \nu + c$ characterizes the first-order slope of the curve, which corresponds to the skew of the implied volatility curve. Meanwhile, the coefficient $\left(\frac{1}{6}-\frac{1}{4}\rho^2\right)\frac{\nu^2}{\alpha}$ mainly controls the second-order curvature of the curve.

To provide a more intuitive illustration of how each parameter affects the shape of the implied volatility surface in the skew-SABR model, we fix a set of baseline parameters
$\alpha = 0.55,\ \rho = -0.2,\ \nu = 2.8,\ c = 0.3,\ d = -0.3$,
and adopt a ceteris paribus approach by varying one parameter at a time while keeping the others fixed. This allows us to isolate and examine the impact of each parameter on the shape of the implied volatility curve.

For market variables such as the forward price, time to maturity, and strike prices, we use the same dataset as in Section~\ref{skew-sec2} for consistency with the analysis of the Hagan-SABR model. Specifically, we take the observed market data of the CSI 1000 equity index option with maturity in June 2025 (contract code MO202506) at 09:50:05 on April 17, 2025 as the reference. The corresponding forward price is $F = 5685.6$, the time to maturity is $\tau = 0.176$, and the strike price range is $[3900,\,7400]$, with discrete strikes sampled at intervals of 100. This part follows the same setting as in Sec~\ref{skew-sec2}.

First, we examine the impact of different $\alpha + d$ values on the implied volatility curve. It can be observed that changes in $\alpha + d$ lead to corresponding shifts in the overall level of the implied volatility curve. This indicates that, in the skew-SABR model, the overall level of the implied volatility curve is jointly determined not only by $\alpha$ but also by the level adjustment term $d$.

\begin{figure}[htbp]
    \centering
\includegraphics[width=0.99\textwidth]{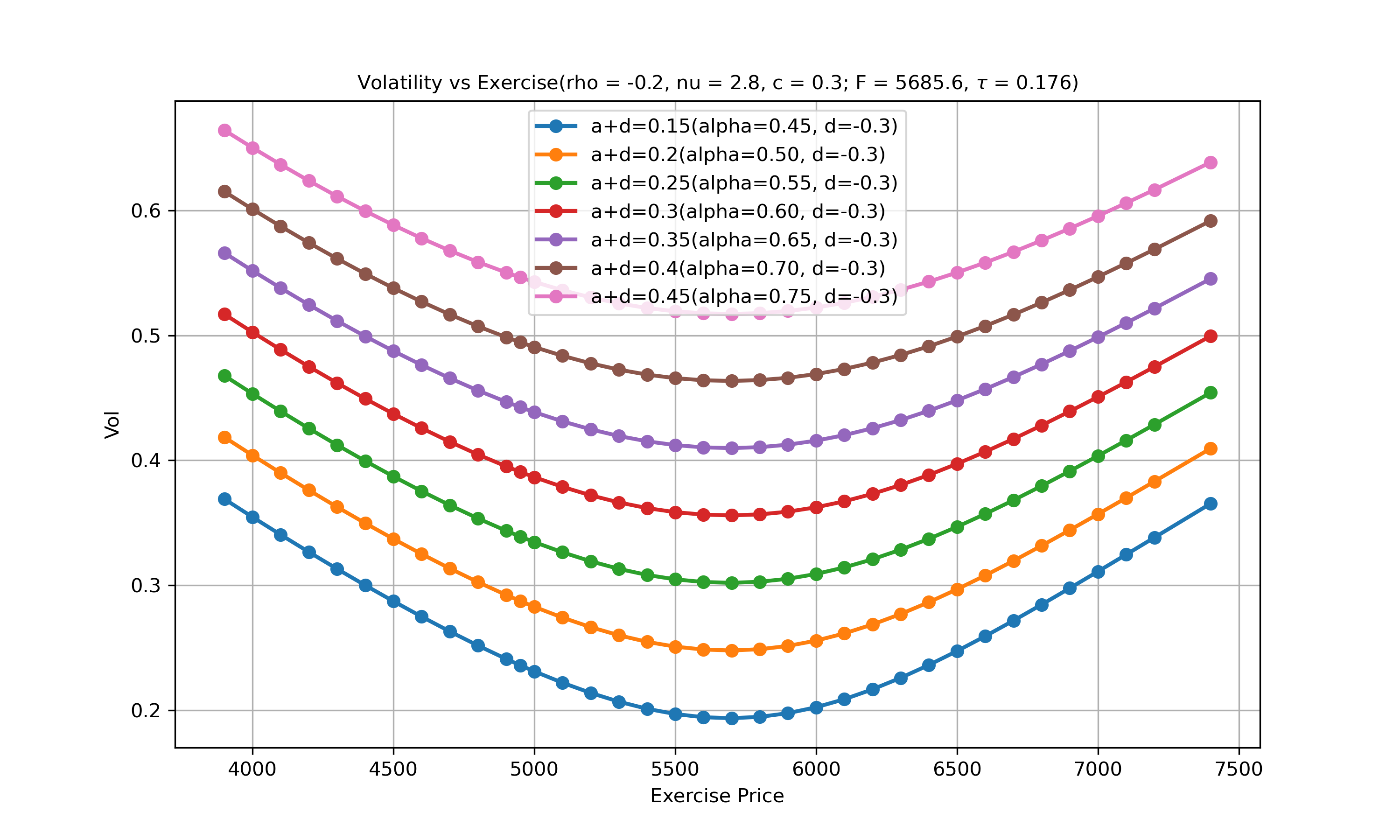}\\
\caption{Implied volatility curves generated by the skew-SABR model under different values of $\alpha + d$}
\end{figure}

Holding $\alpha + d$ constant, we vary $\alpha$ and $d$ separately. The results show that the overall level of the implied volatility curve remains nearly unchanged, while the curvature increases as $\alpha$ decreases. This phenomenon can be explained by the second-order coefficient in the implied volatility expansion, which contains the term $\left(\frac{1}{6}-\frac{1}{4}\rho^2\right)\frac{\nu^2}{\alpha}$. As $\alpha$ decreases, this coefficient increases, leading to a more pronounced curvature of the curve.

\begin{figure}[htbp]
    \centering
\includegraphics[width=0.99\textwidth]{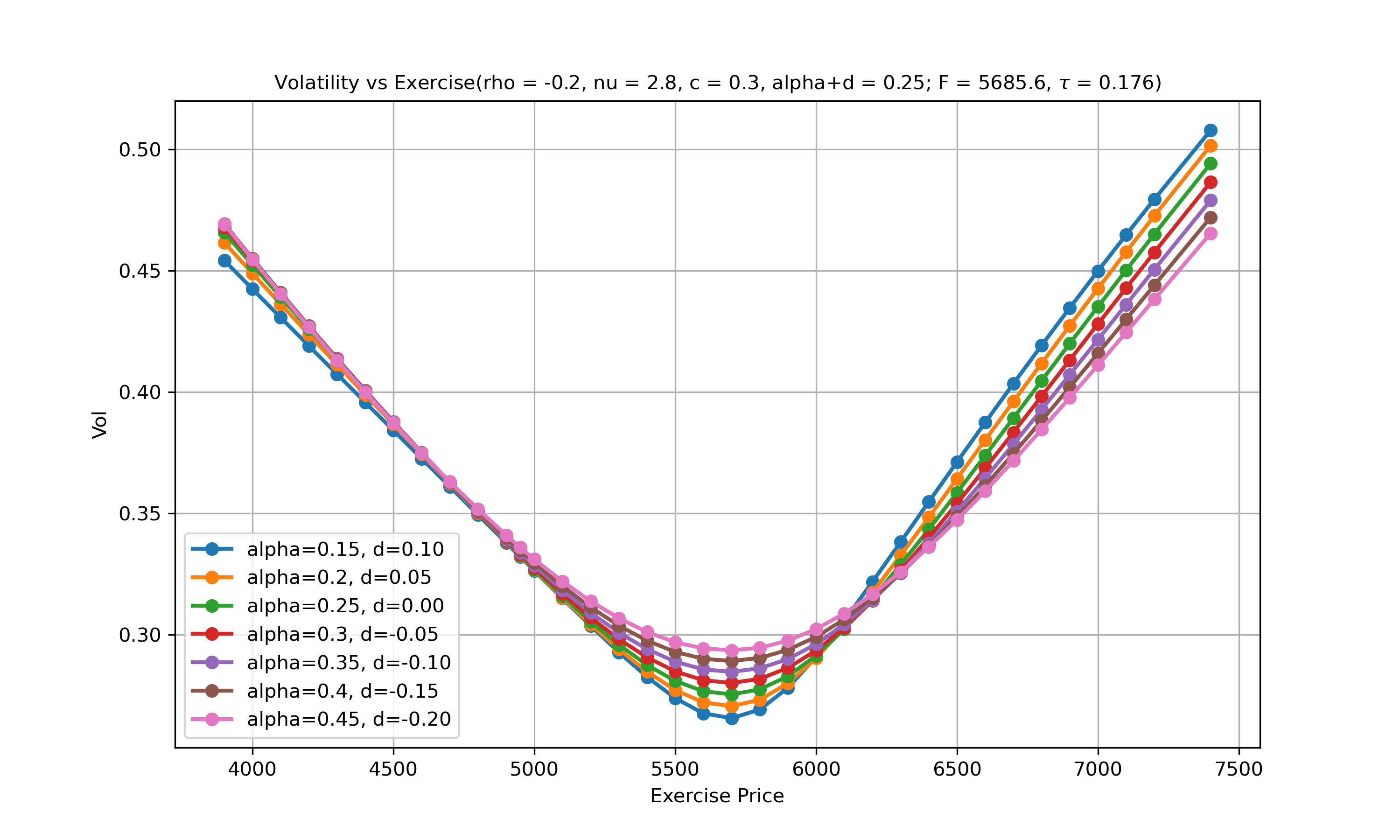}\\
\caption{Implied volatility curves generated by the skew-SABR model for different values of $\alpha$ and $d$  with $\alpha+d$ fixed}
\end{figure}

\begin{figure}[htbp]
    \centering
\includegraphics[width=0.99\textwidth]{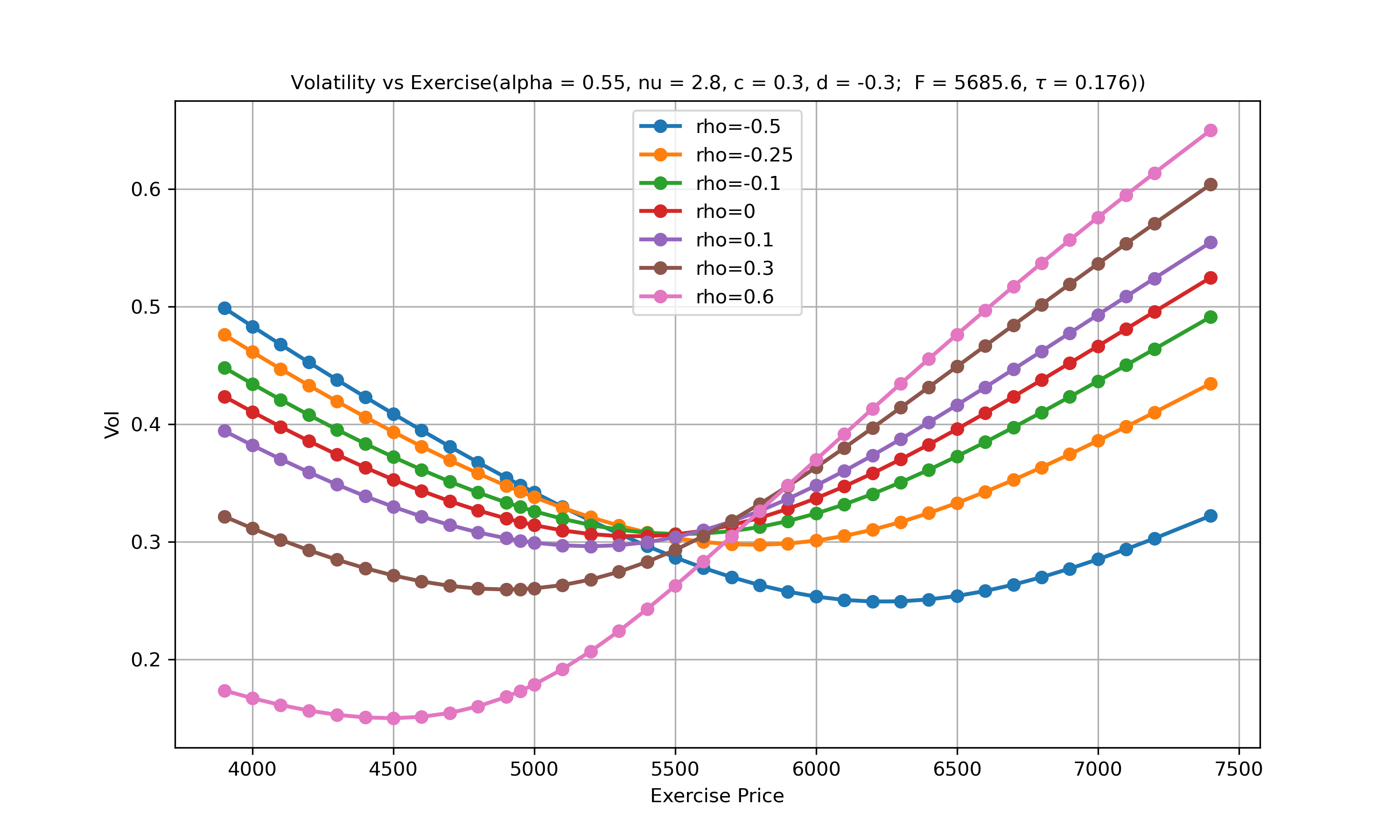}\\
\caption{Implied volatility curves generated by the skew-SABR model under different values of $\rho$} \label{skew-fig6}
\end{figure}

Fig.~\ref{skew-fig6} illustrates the effect of the parameter $\rho$ on the skew of the implied volatility curve. The range of $\rho$ is consistent with that used in Fig.~\ref{skew-fig4}. As shown in the figure, unlike in the Hagan-SABR model, the effect of positive and negative values of $\rho$ on the skew is no longer symmetric in the skew-SABR model. This difference arises from the distinct structure of the skew coefficient in the implied volatility expansion: in the Hagan-SABR model, the coefficient is $\frac{1}{2}\rho \nu$, whereas in the skew-SABR model it becomes $\frac{1}{2}\rho \nu + c$. The introduction of the parameter $c$ implies that the skew is jointly determined by $\rho$ and $c$, thereby breaking the symmetry with respect to the sign of $\rho$.

\begin{figure}[htbp]
    \centering
\includegraphics[width=0.99\textwidth]{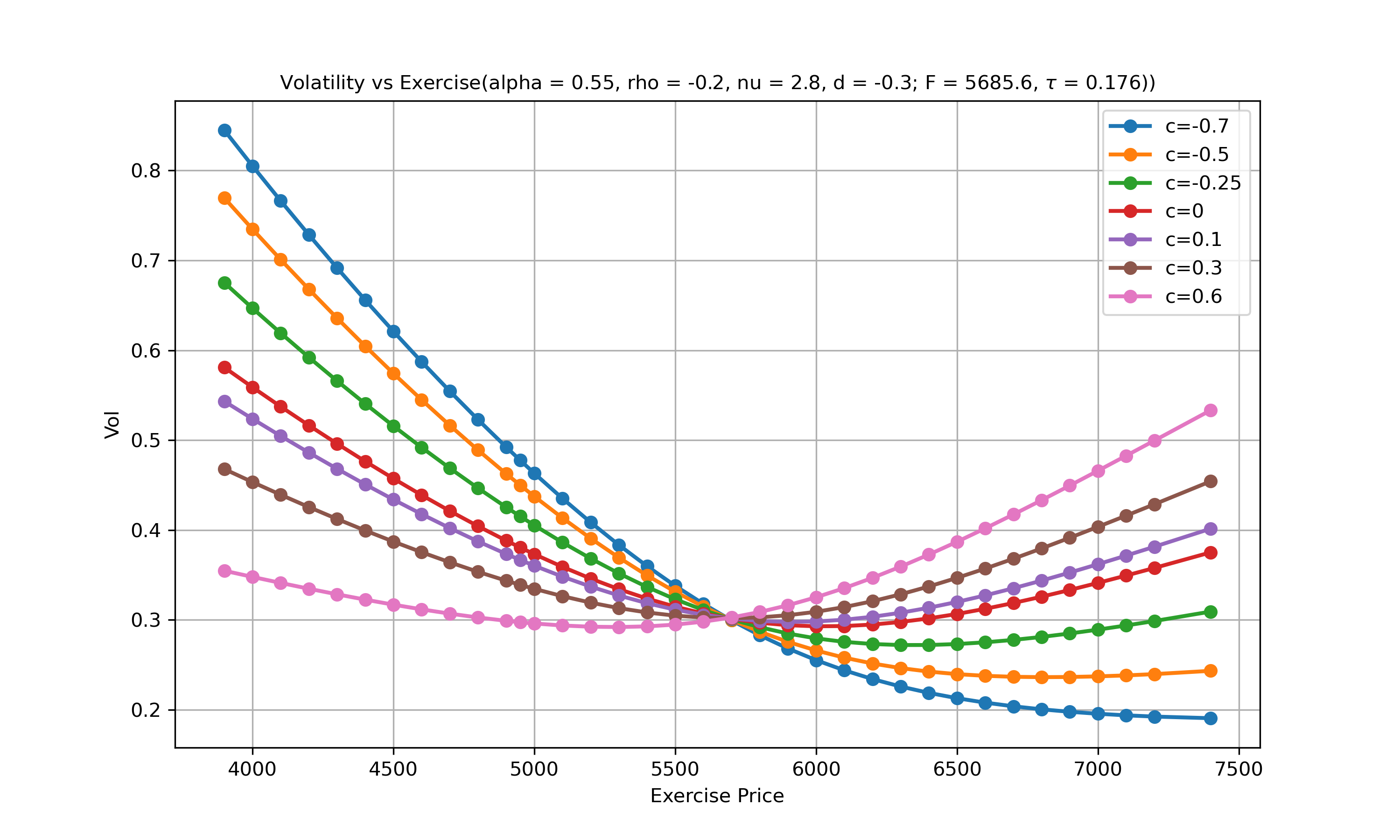}\\
\caption{Implied volatility curves generated by the skew-SABR model under different values of $c$} \label{skew-fig7}
\end{figure}

We further examine the effect of the parameter $c$ on the shape of the implied volatility curve. Fig.~\ref{skew-fig7} presents the implied volatility curves obtained under different values of $c$. The results indicate that $c$ also has a significant impact on the skew of the curve, confirming that $c$ and $\rho$ jointly determine the skew structure in the skew-SABR model.

\begin{figure}[htbp]
    \centering
\includegraphics[width=0.99\textwidth]{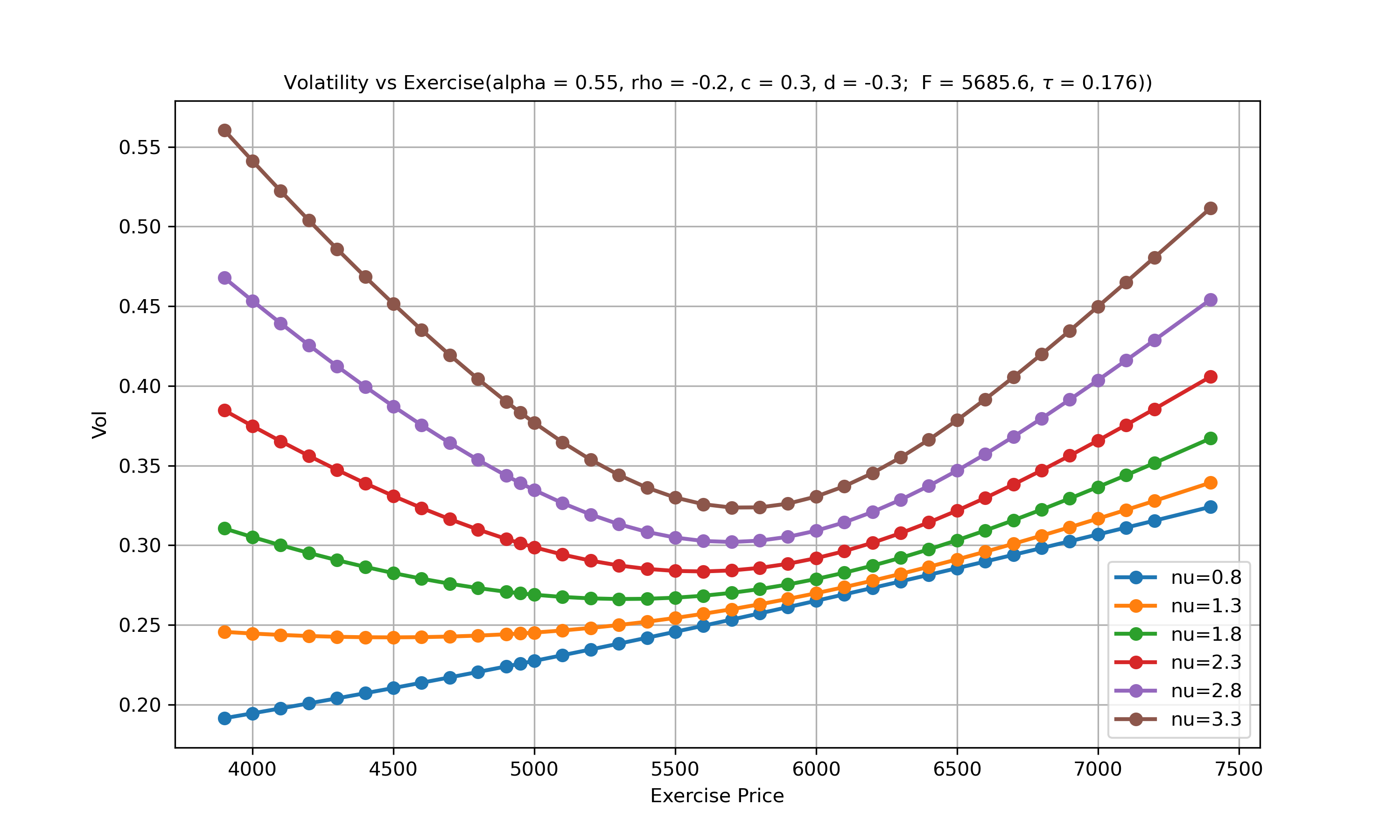}\\
\caption{Implied volatility curves generated by the skew-SABR model under different values of $\nu$} \label{skew-fig8}
\end{figure}

Finally, we analyze the effect of the parameter $\nu$ on the shape of the implied volatility curve. As shown in Fig.~\ref{skew-fig8}, $\nu$ is positively correlated with the curvature: larger values of $\nu$ lead to a more pronounced curvature. This observation is consistent with the theoretical result that the second-order coefficient in the expansion contains the term $\left(\frac{1}{6}-\frac{1}{4}\rho^2\right)\frac{\nu^2}{\alpha}$, indicating that $\nu$ governs the curvature through its influence on the second-order term.

By individually perturbing each parameter and observing the resulting changes in the implied volatility curve, we can obtain a clearer understanding of the geometric roles of the parameters in the skew-SABR model. Specifically, $\alpha + d$ primarily determines the overall level of the curve, $\rho$ and $c$ jointly affect the skew structure, and $\nu$ controls the curvature. These findings are consistent with the analytical results derived from the Taylor expansion of the implied volatility formula. 
Through the combined effect of these parameters, the model is capable of effectively capturing the volatility smile and skew observed in the market, thereby improving its ability to fit real implied volatility curves.

\section{Empirical Results and Model Comparison Analysis}

\label{skew-sec4}

This section evaluates the performance of the proposed skew-SABR model in fitting the market implied volatility surface. To this end, we compare its fitting results with those of the original Hagan-SABR model. Meanwhile, to control for the effect of model complexity on fitting performance, we select three additional models with the same degrees of freedom (i.e., the same number of parameters) as the skew-SABR model as benchmarks, including the SVI (Stochastic Volatility Inspired) model, the polynomial model, and spline interpolation methods. Among these, the SVI and polynomial models are parametric approaches, while spline interpolation represents a nonparametric method. Such a comparison allows for a comprehensive evaluation of the relative performance of the skew-SABR model.

The market data used in this study were obtained from the financial database provided by the International Digital Economy Academy (IDEA), for which the author was granted access during an internship.

\subsection{Description of Benchmark Models}

\textbf{SVI Model\quad} The SVI (Stochastic Volatility Inspired) model originates from the work presented in \cite{gatheral2004parsimonious, gatheral2014arbitrage}, with its early version proposed by Merrill Lynch in 1999. This model is a static parametric implied volatility model, whose core idea is to characterize the total variance structure corresponding to the implied volatility smile by constructing an analytical function for a fixed maturity.

Let $\tau = T - t$ denote the time to maturity, and $k = \log \frac{K}{f}$ denote the log-moneyness. Let the Black implied volatility be denoted by $\sigma_B(k)$. The total variance is defined as
\begin{equation*}
    \omega(k) = \sigma_B^2(k) \cdot \tau.
\end{equation*}

The SVI model fits $\omega(k)$ using the following analytical function:
\begin{equation}
 \omega^{svi}(k) = a + b\big(\rho (k-m) + \sqrt{(k-m)^2 + \sigma^2 } \big),   \label{skew-11}
\end{equation}

As shown in Eq.~(\ref{skew-11}), the model contains five parameters to be calibrated: $a$, $b$, $\rho$, $m$, and $\sigma$, which matches the number of parameters in the skew-SABR model. The parameters are subject to the constraints $a>0$, $b>0$, $|\rho| \leq 1$, and $\sigma > 0$ during calibration.

The calibration procedure is as follows. First, we obtain the market-observed implied volatility $\sigma_{\text{market}}(K, f)$ for different strike prices $K$, and express it as a function of log-moneyness $k$, denoted by $\sigma_{\text{market}}(k)$. Then, the market total variance is computed as $\omega_{\text{market}}(k) = \sigma_{\text{market}}^2(k) \cdot \tau$. Finally, the model parameters are estimated by solving a nonlinear least squares problem that minimizes the weighted squared error between the model-implied and market-observed total variance:
\begin{equation*}
   a, b, \rho, m, \sigma = \arg \min \sum_k w_k \left|\omega^{svi}(k)- \omega_{\text{market}}(k)\right|^2.
\end{equation*}
Here, $w_k$ denotes the weight associated with each strike. After calibration, the fitted implied volatility curve can be recovered via $\sigma_B^{svi}(k) = \sqrt{\omega^{svi}(k) / \tau}$.

From a modeling perspective, the SVI model is a purely parametric approach, primarily motivated by empirical observations of market data rather than derived from a rigorous theoretical framework. Therefore, its economic interpretability is relatively limited, and the conomic implications of its parameters are typically
need to be interpreted in conjunction with their mathematical expressions: parameter $a$ mainly determines the overall level of the total variance curve; $b$ controls the slope of the wings; $\rho$ governs the asymmetry of the curve, analogous to the conventional notion of skew; $m$ represents the shift of the log-strike at which the implied volatility attains its minimum relative to the at-the-money point; and $\sigma$ controls the curvature and smoothness of the curve in the central region.

\textbf{Polynomial Model\quad}
The polynomial model is one of the simplest parametric approaches for modeling implied volatility. Its basic idea is to directly assume that the implied volatility is a polynomial function of the log-moneyness, i.e.,
\begin{equation*}
    \sigma_B^{poly}(k) = \sum_{i=0}^{n} a_i k^i,
\end{equation*}
where $k = \log \frac{K}{f}$ denotes the log-moneyness, $K$ is the strike price, and $f$ is the forward price.

In the numerical experiments of this study, we set $n=4$, corresponding to a quartic polynomial. The model therefore contains five parameters $\{a_i\}_{i=0}^{4}$, which is consistent with the number of parameters in the skew-SABR model. The parameters are estimated via a least-squares approach by solving
\begin{equation*}
\min_{a_i} \sum w_k\left(\sigma_{\text{market}}(k) - \sigma_B^{poly}(k)\right)^2.
\end{equation*}

\textbf{Spline Method\quad}
Spline functions are a class of piecewise polynomial methods widely used for fitting nonlinear functions. In this study, B-splines are employed to approximate the implied volatility curve, and the parameters are estimated using the \texttt{LSQUnivariateSpline} algorithm provided in the SciPy library. Since this algorithm performs least-squares fitting directly based on the input values of the independent variable, the strike price $K$ is used as the independent variable.

Let $\sigma_B^{spline}(K)$ denote the implied volatility curve represented by the B-spline function, which can be written as
\begin{equation*}
    \sigma_B^{spline}(K) = \sum_{i=0}^{n-1} c_i B_{i,l}(K),
\end{equation*}
where $B_{i,l}(K)$ denotes the B-spline basis functions of degree $l$, and $c_i$ are the coefficients to be estimated.

The model parameters are calibrated using a least-squares method. Here, an equal-weight least-squares approach is adopted, i.e.,
\begin{equation*}
\min_{c_i} \sum \left(\sigma_{\text{market}}(K) - \sigma_B^{spline}(K)\right)^2.
\end{equation*}

Let $l$ denote the degree of the spline and $m$ the number of interior knots. Then, the number of B-spline basis functions is given by $n = l + m + 1$. In this study, a cubic spline is employed, i.e., $l=3$, with one interior knot ($m=1$). The interior knot is chosen as the strike price closest to the forward price $F$; if this strike lies on the boundary of the interval, the interior strike closest to $F$ is selected instead.

\subsection{Empirical Results and Model Comparison}

This study considers various options, including 
CSI 1000 equity index (MO) options, CSI 300 equity index (IO) options, SSE 50 equity index (HO) options, CSI 300 ETF (510300) options, CSI 500 ETF (510500) options, SSE 50 ETF (510050) options, and E Fund ChiNext ETF (159915) options, to systematically evaluate the fitting performance of the models. To comprehensively assess model performance under different market conditions and time scales, the empirical analysis is divided into two stages.

The first stage focuses on the first half of 2025 (denoted by H1 2025 in the table titles below). In this stage, high-frequency data are employed, and all listed option contracts during this period are evaluated using 5-minute frequency data. This setup ensures coverage of all trading timestamps throughout the entire lifecycle of each contract. Such high-frequency data with full lifecycle coverage enables a detailed evaluation of model performance across different maturities and market conditions, as well as an assessment of model stability over the entire lifespan of individual contracts.

The second stage spans from 2018 to the end of 2024. This period largely covers the historical data since the listing of major financial option products in China, and thus provides a basis for evaluating model performance over a long time horizon. Due to the extended time span and large data volume, computational complexity is controlled by selecting only one option contract per underlying at each timestamp and performing model calibration at four fixed time points per trading day. This design allows for an efficient evaluation of overall model performance on long-term historical data.

Combining the two stages, the sample period used in this study ranges from January 1, 2018 to June 30, 2025.

In the first-stage analysis for the first half of 2025, option contracts with maturities in April, May, June, and September 2025 are selected for comparative analysis. For the April, May, and June contracts, the evaluation period runs from the listing date to the day prior to expiration. For the September contracts, since they do not expire within the sample period, the evaluation period spans from the listing date to June 30, 2025.

Within each trading day, model parameters are calibrated every 5 minutes from 9:30:05 to 14:55:05, and the corresponding fitting errors are computed. The root mean squared error (RMSE) is adopted as the evaluation metric for fitting accuracy. Suppose that at a given timestamp there are $n$ available strike prices for implied volatility calculation, then the RMSE is defined as
\begin{equation*}
    \mathrm{RMSE} = \sqrt{\frac{1}{n}\sum_{i=1}^{n} \left(\sigma_{\text{model}}^i - \sigma_{\text{market}}^i\right)^2 } .
\end{equation*}

Subsequently, the RMSE values obtained at each timestamp are aggregated, and the average RMSE is computed for each model across different underlyings and contracts.

First, we present the results for   equity index options, including MO, IO, and HO. Each row in the table corresponds to one of the five models: Hagan-SABR, skew-SABR, SVI, cubic spline, and polynomial; each column corresponds to a specific contract, with the associated sample period reported in parentheses. For the April, May, and June contracts, the sample period spans from the listing date to the day prior to expiration, whereas for the September contracts, the sample period runs from the listing date to June 30, 2025. The listing period of non-quarterly contracts is approximately three months, while that of quarterly contracts is about one year. The values highlighted in bold indicate the model that achieves the lowest RMSE for the corresponding contract.

\begin{table}[H]
    \centering
 \caption{Model fitting results for CSI 1000 equity index (MO) options (H1 2025)}
\renewcommand{\arraystretch}{1.5}   
\tabcolsep=0.2cm  

    \begin{tabular}{|c|c|c|c|c|}
    \hline
        MO & \makecell{MO202504 \\ (250120-250417)}
 & \makecell{MO202505 \\ (250224-250515)} &  \makecell{MO202506 \\ (240624-250623)}  & \makecell{MO202509 \\ (240923-250630)} \\
\hline
        Hagan SABR & 7.087$\times 10^{-3}$ & 6.779$\times 10^{-3}$ & 5.791$\times 10^{-3}$ & 2.177$\times 10^{-3}$  \\ \hline
        skew SABR & 2.723$\times 10^{-3}$ & \textbf{2.062}$\times\textbf{10}^{\textbf{-3}}$ & \textbf{2.294}$\times \textbf{10}^{\textbf{-3}}$ & 9.724$\times 10^{-4}$  \\ \hline
        SVI & 7.004$\times 10^{-3}$ & 8.165$\times 10^{-3}$ & 6.245$\times 10^{-3}$ & 1.589$\times 10^{-3}$  \\ \hline
        cubic spline & \textbf{2.469$\times \textbf{10}^{\textbf{-3}}$} & 2.185$\times 10^{-3}$ & 2.424$\times 10^{-3}$ & \textbf{8.721$\times \textbf{10}^{\textbf{-4}}$}  \\ \hline
        polynomial & 2.897$\times 10^{-3}$ & 3.016$\times 10^{-3}$ & 6.995$\times 10^{-3}$ & 1.502$\times 10^{-3}$  \\ \hline \end{tabular}
\end{table}

\begin{table}[H]
    \centering
 \caption{Model fitting results for CSI 300 equity index (IO) options (H1 2025)}
\renewcommand{\arraystretch}{1.5}   
\tabcolsep=0.2cm  

    \begin{tabular}{|c|c|c|c|c|}
    \hline
        IO & \makecell{IO202504 \\ (250120-250417)}
 & \makecell{IO202505 \\ (250224-250515)} &  \makecell{IO202506 \\ (240624-250623)}  & \makecell{IO202509 \\ (240923-250630)} \\
\hline
     Hagan SABR & 6.268$\times 10^{-3}$ & 4.575$\times 10^{-3}$ & 4.012$\times 10^{-3}$ & 1.328$\times 10^{-3}$  \\ \hline
    skew SABR & \textbf{1.859$\times \textbf{10}^{\textbf{-3}}$} & \textbf{2.023$\times \textbf{10}^{\textbf{-3}}$} & \textbf{1.299$\times \textbf{10}^{\textbf{-3}}$} & 9.109$\times 10^{-4}$  \\ \hline
    SVI & 7.314$\times 10^{-3}$ & 8.898$\times 10^{-3}$ & 5.992$\times 10^{-3}$ & 1.328$\times 10^{-3}$  \\ \hline
    cubic spline & 2.302$\times 10^{-3}$ & 2.961$\times 10^{-3}$ & 1.764$\times 10^{-3}$ & \textbf{7.488$\times \textbf{10}^{\textbf{-4}}$}  \\ \hline
    polynomial & 2.513$\times 10^{-3}$ & 4.212$\times 10^{-3}$ & 4.075$\times 10^{-3}$ & 1.427$\times 10^{-3}$  \\ \hline \end{tabular}
\end{table}

\begin{table}[H]
    \centering
 \caption{Model fitting results for SSE 50 equity index (HO) options (H1 2025)}
\renewcommand{\arraystretch}{1.5}   
\tabcolsep=0.2cm  

    \begin{tabular}{|c|c|c|c|c|}
    \hline
        HO & \makecell{HO202504 \\ (250120-250417)}
 & \makecell{HO202505 \\ (250224-250515)} &  \makecell{HO202506 \\ (240624-250623)}  & \makecell{HO202509 \\ (240923-250630)} \\
\hline
    Hagan SABR & 5.478$\times 10^{-3}$ & 4.993$\times 10^{-3}$ & 4.367$\times 10^{-3}$ & 1.752$\times 10^{-3}$  \\ \hline
    skew SABR & \textbf{2.623$\times \textbf{10}^{\textbf{-3}}$} & \textbf{2.713$\times \textbf{10}^{\textbf{-3}}$} & \textbf{2.034$\times \textbf{10}^{\textbf{-3}}$} & 1.186$\times 10^{-3}$  \\ \hline
    SVI & 7.230$\times 10^{-3}$ & 8.359$\times 10^{-3}$ & 5.712$\times 10^{-3}$ & 1.615$\times 10^{-3}$  \\ \hline
    cubic spline & 2.904$\times 10^{-3}$ & 3.123$\times 10^{-3}$ & 2.331$\times 10^{-3}$ & \textbf{1.145$\times \textbf{10}^{\textbf{-3}}$}  \\ \hline
    polynomial & 3.126$\times 10^{-3}$ & 3.962$\times 10^{-3}$ & 4.113$\times 10^{-3}$ & 1.480$\times 10^{-3}$  \\ \hline  \end{tabular}
\end{table}

We next present the results for ETF options. Compared with equity index options, ETF option contracts typically have shorter listing periods. The RMSE statistics of different models across contracts are reported as follows.

\begin{table}[H]
    \centering
 \caption{Model fitting results for CSI 500 ETF (510500) options (H1 2025)}
\renewcommand{\arraystretch}{1.5}   
\tabcolsep=0.2cm  

    \begin{tabular}{|c|c|c|c|c|}
    \hline
        510500 & \makecell{510500$\_$202504 \\ (250227-250422)}
 & \makecell{510500$\_$202505 \\ (250327-250527)} &  \makecell{510500$\_$202506 \\ (241024-250624)}  & \makecell{510500$\_$202509 \\ (250123-250630)} \\
\hline
        Hagan-SABR & 7.205$\times 10^{-3}$ & 6.866$\times 10^{-3}$ & 3.162$\times 10^{-3}$ & 1.018$\times 10^{-3}$  \\ \hline
        \textbf{skew-SABR} & \textbf{3.293$\times \textbf{10}^{\textbf{-3}}$} & \textbf{3.441$\times \textbf{10}^{\textbf{-3}}$} & \textbf{1.321$\times \textbf{10}^{\textbf{-3}}$} & 5.515$\times 10^{-4}$  \\ \hline
        SVI & 1.103$\times 10^{-2}$ & 1.141$\times 10^{-2}$ & 3.886$\times 10^{-3}$ & 8.189$\times 10^{-4}$  \\ \hline
        cubic spline& 5.334$\times 10^{-3}$ & 4.275$\times 10^{-3}$ & 1.615$\times 10^{-3}$ & \textbf{4.568$\times 10^{-4}$}  \\ \hline
        polynomial & 1.283$\times 10^{-2}$ & 5.539$\times 10^{-3}$ & 2.509$\times 10^{-3}$ & 1.086$\times 10^{-3}$  \\ \hline   \end{tabular}
\end{table}

\begin{table}[H]
    \centering
 \caption{Model fitting results for CSI 300 ETF (510300) options (H1 2025)}
\renewcommand{\arraystretch}{1.5}   
\tabcolsep=0.2cm  

    \begin{tabular}{|c|c|c|c|c|}
    \hline
        510300 & \makecell{510300$\_$202504 \\ (250227-250422)}
 & \makecell{510300$\_$202505 \\ (250327-250527)} &  \makecell{510300$\_$202506 \\ (241024-250624)}  & \makecell{510300$\_$202509 \\ (250123-250630)} \\
\hline
        Hagan SABR & 5.707$\times 10^{-3}$ & 5.788$\times 10^{-3}$ & 2.164$\times 10^{-3}$ & 3.978$\times 10^{-4}$  \\ \hline
        skew SABR & \textbf{2.390$\times \textbf{10}^{\textbf{-3}}$} & \textbf{2.491$\times \textbf{10}^{\textbf{-3}}$} & 8.429$\times 10^{-4}$ & 3.081$\times 10^{-4}$  \\ \hline
        SVI & 8.790$\times 10^{-3}$ & 1.041$\times 10^{-2}$ & 3.300$\times 10^{-3}$ & 4.016$\times 10^{-4}$  \\ \hline
        cubic spline & 2.884$\times 10^{-3}$ & 2.679$\times 10^{-3}$ & \textbf{7.604$\times \textbf{10}^{\textbf{-4}}$} & \textbf{2.579$\times \textbf{10}^{\textbf{-4}}$}  \\ \hline
        polynomial & 3.240$\times 10^{-3}$ & 3.645$\times 10^{-3}$ & 9.217$\times 10^{-4}$ & 2.941$\times 10^{-4}$  \\ \hline
    \end{tabular}
\end{table}

\begin{table}[H]
    \centering
 \caption{Model fitting results for SSE 50 ETF (510050) options (H1 2025)}
\renewcommand{\arraystretch}{1.5}   
\tabcolsep=0.2cm  

    \begin{tabular}{|c|c|c|c|c|}
    \hline
        510050 & \makecell{510050$\_$202504 \\ (250227-250422)}
 & \makecell{510050$\_$202505 \\ (250327-250527)} &  \makecell{510050$\_$202506 \\ (241024-250624)}  & \makecell{510050$\_$202509 \\ (250123-250630)} \\
\hline
 Hagan SABR & 4.153$\times 10^{-3}$ & 6.491$\times 10^{-3}$ & 2.138$\times 10^{-3}$ & 5.615$\times 10^{-4}$  \\ \hline
skew SABR & \textbf{1.761$\times \textbf{10}^{\textbf{-3}}$} & 2.027$\times 10^{-3}$ & \textbf{6.974$\times \textbf{10}^{\textbf{-4}}$} & 4.615$\times 10^{-4}$  \\ \hline
SVI & 5.746$\times 10^{-3}$ & 7.872$\times 10^{-3}$ & 2.474$\times 10^{-3}$ & 5.380$\times 10^{-4}$  \\ \hline
cubic spline & 1.859$\times 10^{-3}$ & \textbf{1.980$\times \textbf{10}^{\textbf{-3}}$} & 7.583$\times 10^{-4}$ & \textbf{3.890$\times \textbf{10}^{\textbf{-4}}$}  \\ \hline
polynomial & 2.116$\times 10^{-3}$ & 2.411$\times 10^{-3}$ & 9.909$\times 10^{-4}$ & 4.054$\times 10^{-4}$  \\ \hline  \end{tabular}
\end{table}

\begin{table}[H]
    \centering
 \caption{Model fitting results for ChiNext ETF (159915) options (H1 2025)}
\renewcommand{\arraystretch}{1.5}   
\tabcolsep=0.2cm  

    \begin{tabular}{|c|c|c|c|c|}
    \hline
        159915 & \makecell{159915$\_$202504 \\ (250227-250422)}
 & \makecell{159915$\_$202505 \\ (250327-250527)} &  \makecell{159915$\_$202506 \\ (241024-250624)}  & \makecell{159915$\_$202509 \\ (250123-250630)} \\
\hline
Hagan SABR & 5.454$\times 10^{-3}$ & 8.585$\times 10^{-3}$ & 3.304$\times 10^{-3}$ & 8.029$\times 10^{-4}$  \\ \hline
skew SABR & \textbf{3.250$\times \textbf{10}^{\textbf{-3}}$} & \textbf{2.791$\times \textbf{10}^{\textbf{-3}}$} & \textbf{1.316$\times \textbf{10}^{\textbf{-3}}$} & 6.549$\times 10^{-4}$  \\ \hline
SVI & 7.730$\times 10^{-3}$ & 1.033$\times 10^{-2}$ & 4.441$\times 10^{-3}$ & 6.996$\times 10^{-4}$  \\ \hline
cubic spline & 4.465$\times 10^{-3}$ & 2.885$\times 10^{-3}$ & 1.709$\times 10^{-3}$ & \textbf{5.528$\times \textbf{10}^{\textbf{-4}}$}  \\ \hline
polynomial & 6.446$\times 10^{-3}$ & 3.347$\times 10^{-3}$ & 2.614$\times 10^{-3}$ & 7.189$\times 10^{-4}$  \\ \hline  \end{tabular}
\end{table}

From the above results, it can be observed that the experiments in the first half of 2025 provide a detailed evaluation of model performance under recent market conditions. The analysis considers two quarterly contracts and two non-quarterly contracts for each underlying. Overall, the skew-SABR model achieves the lowest fitting error in most cases, while in a few instances the cubic spline method attains slightly better accuracy. It should be noted that the cubic spline method is a purely data-driven interpolation approach; although it can achieve high fitting accuracy, it lacks a clear financial interpretation.

In addition, a relatively robust pattern can be observed across different contracts: the skew-SABR model tends to exhibit superior fitting performance for near-maturity options. Since the April, May, and June contracts experience their full lifecycle within the sample period, including a large number of near-expiry trading intervals, the skew-SABR model typically achieves lower fitting errors for these contracts. In contrast, the September contracts remain in a longer-dated regime during the sample period, under which the cubic spline method may occasionally produce slightly lower fitting errors.

Beyond the detailed analysis of recent market conditions, this study further evaluates model performance over a longer time horizon. Specifically, for each option type, data are collected from the second month after its listing date through the end of 2024. Since SSE 50 ETF options were listed earlier than other products, their sample period is set from January 2018 to December 2024. The listing dates and corresponding sample periods for each option type are summarized in Table~\ref{skew-table9}.

\begin{table}[H]
    \centering
 \caption{Listing dates and sample periods for each option type}
\renewcommand{\arraystretch}{1.5}
\tabcolsep=0.2cm
    
    \begin{tabular}{|c|c|c|}
    \hline
        Option Type & Listing Date & Sample Period  \\ \hline
        MO & 2022.7.22 & 2022.8--2024.12  \\ \hline
        IO & 2019.12.23 & 2020.1--2024.12  \\ \hline
        HO & 2022.12.19 & 2023.1--2024.12  \\ \hline
        510500 & 2022.9.19 & 2022.10--2024.12  \\ \hline
        510300 & 2019.12.23 & 2020.1--2024.12  \\ \hline
        510050 & 2015.2.9 & 2018.1--2024.12  \\ \hline
        159915 & 2022.9.19 & 2022.10--2024.12  \\ \hline 
    \end{tabular}  \label{skew-table9}
\end{table}

The above sample periods largely cover the full listing history of major financial option products in China, thereby providing a comprehensive assessment of model performance under long-term market conditions. Due to the large data volume in this stage, only one contract is selected per trading day for computation. Specifically, the near-month option is chosen for fitting, where the near-month contract is defined as the next-to-expire quarterly contract. In addition, data on the expiration date of each contract are excluded.

Taking MO options as an example, this product was listed on July 22, 2022, and the sample period starts from August 1, 2022. Since the MO2209 contract expired on September 16, 2022, it is used for computation from August 1 to September 15, 2022; thereafter, starting from September 16, 2022, the MO2212 contract is used. This rolling selection ensures that the most representative near-month contract is consistently chosen throughout the sample period.

Furthermore, to control computational costs, only four time points are selected per trading day for model calibration, namely 9:45, 10:45, 13:45, and 14:45. These time points correspond to the market open, mid-morning session, mid-afternoon session, and the period close to market close, respectively, thereby capturing different intraday market conditions. Based on the above sampling scheme, the fitting errors of each model are evaluated across different option types and years.

First, we report the statistical results for index options (MO, IO, HO).

\begin{table}[H]
    \centering
 \caption{Model fitting results for CSI 1000 equity index (MO) options (Aug 2022 -- Dec 2024)}
\renewcommand{\arraystretch}{1.5}
\tabcolsep=0.2cm
    \begin{tabular}{|c|c|c|c|c|}
    \hline \makecell{MO \\ (2022.8-2024.12)} & 2022 & 2023 & 2024 & all  \\ \hline
        Hagan SABR & 8.377$\times 10^{-3}$ & 1.229$\times 10^{-2}$ & 2.205$\times 10^{-2}$ & 1.561$\times 10^{-2}$  \\ \hline
        skew SABR & \textbf{2.231$\times \textbf{10}^{\textbf{-3}}$} & \textbf{3.515$\times \textbf{10}^{\textbf{-3}}$} & 7.786$\times 10^{-3}$ & 5.046$\times 10^{-3}$  \\ \hline
        SVI & 9.481$\times 10^{-3}$ & 1.458$\times 10^{-2}$ & 2.161$\times 10^{-2}$ & 1.657$\times 10^{-2}$  \\ \hline
        cubic spline & 2.427$\times 10^{-3}$ & 3.633$\times 10^{-3}$ & \textbf{6.558$\times \textbf{10}^{\textbf{-3}}$} & \textbf{4.624$\times \textbf{10}^{\textbf{-3}}$ } \\ \hline
        polynomial & 2.716$\times 10^{-3}$ & 5.407$\times 10^{-3}$ & 4.441$\times 10^{-2}$ & 2.098$\times 10^{-2}$  \\ \hline
    \end{tabular}
\end{table}

\begin{table}[H]
    \centering
 \caption{Model fitting results for CSI 300 equity index (IO) options (Jan 2020 -- Dec 2024)}
\renewcommand{\arraystretch}{1.5}
\tabcolsep=0.2cm
    \begin{tabular}{|c|c|c|c|c|}
 \hline \makecell{IO \\ (2020.1-2024.12)} & 2020 & 2021 & 2022  \\ \hline
        Hagan SABR & 1.498$\times 10^{-2}$ & 1.927$\times 10^{-2}$ & 1.791$\times 10^{-2}$   \\ \hline
        skew SABR & \textbf{2.991}$\times \textbf{10}^{\textbf{-3}}$ & \textbf{3.459$\times \textbf{10}^{\textbf{-3}}$} & \textbf{3.792$\times \textbf{10}^{\textbf{-3}}$}   \\ \hline
        SVI & 9.703$\times 10^{-3}$ & 1.453$\times 10^{-2}$ & 1.529$\times 10^{-2}$   \\ \hline
        cubic spline & 3.662$\times 10^{-3}$ & 4.103$\times 10^{-3}$ & 4.114$\times 10^{-3}$   \\ \hline
        polynomial & 9.507$\times 10^{-3}$ & 1.431$\times 10^{-2}$ & 1.676$\times 10^{-2}$   \\ \hline   & 2023 & 2024 & all  \\ \hline
        Hagan SABR & 1.298$\times 10^{-2}$ & 1.275$\times 10^{-2}$ & 1.558$\times 10^{-2}$   \\ \hline
        skew SABR & \textbf{3.245$\times \textbf{10}^{\textbf{-3}}$} & \textbf{3.570$\times \textbf{10}^{\textbf{-3}}$} & \textbf{3.411$\times \textbf{10}^{\textbf{-3}}$}   \\ \hline
        SVI & 1.495$\times 10^{-2}$ & 1.492$\times 10^{-2}$ & 1.388$\times 10^{-2}$   \\ \hline
        cubic spline & 3.994$\times 10^{-3}$ & 4.890$\times 10^{-3}$ & 4.152$\times 10^{-3}$   \\ \hline
        polynomial & 6.929$\times 10^{-3}$ & 8.571$\times 10^{-3}$ & 1.122$\times 10^{-2}$
   \\ \hline
    \end{tabular}
\end{table}

\begin{table}[H]
    \centering
 \caption{Model fitting results for SSE 50 equity index (HO) options (Jan 2023 -- Dec 2024)}
\renewcommand{\arraystretch}{1.5}
\tabcolsep=0.2cm
    \begin{tabular}{|c|c|c|c|c|}
 \hline \makecell{HO \\ (2023.1-2024.12)} & 2023 & 2024 & all  \\ \hline
        Hagan SABR & 1.164$\times 10^{-2}$ & 1.476$\times 10^{-2}$ & 1.320$\times 10^{-2}$  \\ \hline
        skew SABR & \textbf{3.931$\times \textbf{10}^{\textbf{-3}}$} & \textbf{6.523$\times \textbf{10}^{\textbf{-3}}$} & \textbf{5.227$\times \textbf{10}^{\textbf{-3}}$}  \\ \hline
        SVI & 1.116$\times 10^{-2}$ & 1.611$\times 10^{-2}$ & 1.364$\times 10^{-2}$  \\ \hline
        cubic spline & 4.154$\times 10^{-3}$ & 7.072$\times 10^{-3}$ & 5.613$\times 10^{-3}$  \\ \hline
        polynomial & 6.778$\times 10^{-3}$ & 8.233$\times 10^{-3}$ & 7.505$\times 10^{-3}$  \\ \hline
    \end{tabular}
\end{table}

Subsequently, we present the statistical results for ETF options (510300, 510500, 510050, 159915).

\begin{table}[H]
    \centering
 \caption{Model fitting results for CSI 500 ETF (510500) options (Oct 2022 -- Dec 2024)}
\renewcommand{\arraystretch}{1.5}
\tabcolsep=0.2cm
    \begin{tabular}{|c|c|c|c|c|}
\hline \makecell{510500 \\ (2022.10-2024.12)}
     & 2022 & 2023 & 2024 & all  \\ \hline
    Hagan SABR & 5.877$\times 10^{-3}$ & 9.691$\times 10^{-3}$ & 1.299$\times 10^{-2}$ & 9.978$\times 10^{-3}$  \\ \hline
    skew SABR & \textbf{1.796$\times \textbf{10}^{\textbf{-3}}$} & \textbf{2.886$\times \textbf{10}^{\textbf{-3}}$} & 3.418$\times 10^{-3}$ & \textbf{2.962$\times \textbf{10}^{\textbf{-3}}$}  \\ \hline
    SVI & 9.314$\times 10^{-3}$ & 1.158$\times 10^{-2}$ & 1.153$\times 10^{-2}$ & 1.095$\times 10^{-2}$  \\ \hline
    cubic spline & 2.166$\times 10^{-3}$ & 3.236$\times 10^{-3}$ & \textbf{3.404$\times \textbf{10}^{\textbf{-3}}$} & 3.155$\times 10^{-3}$  \\ \hline
    polynomial & 2.561$\times 10^{-3}$ & 5.499$\times 10^{-3}$ & 9.958$\times 10^{-3}$ & 6.698$\times 10^{-3}$  \\ \hline
    \end{tabular}
\end{table}

\begin{table}[H]
    \centering
 \caption{Model fitting results for CSI 300 ETF (510300) options (Jan 2020 -- Dec 2024)}
\renewcommand{\arraystretch}{1.5}
\tabcolsep=0.2cm
    \begin{tabular}{|c|c|c|c|c|}
    \hline \makecell{510300 \\ (2020.1-2024.12)} & 2020 & 2021 & 2022   \\ \hline
        Hagan SABR & 1.318$\times 10^{-2}$ & 9.424$\times 10^{-3}$ & 1.053$\times 10^{-2}$    \\ \hline
        skew SABR & \textbf{3.239$\times \textbf{10}^{\textbf{-3}}$} & 3.273$\times 10^{-3}$ & \textbf{2.661$\times \textbf{10}^{\textbf{-3}}$}    \\ \hline
        SVI & 1.023$\times 10^{-2}$ & 7.201$\times 10^{-3}$ & 9.222$\times 10^{-3}$    \\ \hline
        cubic spline & 3.685$\times 10^{-3}$ & \textbf{2.639$\times \textbf{10}^{\textbf{-3}}$} & 2.775$\times 10^{-3}$    \\ \hline
        polynomial & 7.784$\times 10^{-3}$ & 5.631$\times 10^{-3}$ & 4.901$\times 10^{-3}$   \\ \hline   & 2023 & 2024 & all   \\ \hline
        Hagan SABR & 5.981$\times 10^{-3}$ & 8.630$\times 10^{-3}$ & 9.525$\times 10^{-3}$    \\ \hline
        skew SABR & \textbf{2.090$\times \textbf{10}^{\textbf{-3}}$} & \textbf{2.941$\times \textbf{10}^{\textbf{-3}}$} & \textbf{2.815$\times \textbf{10}^{\textbf{-3}}$}   \\ \hline
        SVI & 5.280$\times 10^{-3}$ & 9.314$\times 10^{-3}$ & 8.548$\times 10^{-3}$    \\ \hline
        cubic spline & 2.270$\times 10^{-3}$ & 3.250$\times 10^{-3}$ & 2.915$\times 10^{-3}$    \\ \hline
        polynomial & 2.642$\times 10^{-3}$ & 4.373$\times 10^{-3}$ & 5.062$\times 10^{-3}$  \\ \hline
    \end{tabular}
\end{table}

\begin{table}[H]
    \centering
 \caption{Model fitting results for SSE 50 ETF (510050) options (Jan 2018 -- Dec 2024)}
\renewcommand{\arraystretch}{1.5}
\tabcolsep=0.2cm
    \begin{tabular}{|c|c|c|c|c|}
\hline
        \makecell{510050 \\ (2018.1-2024.12)} & 2018 & 2019 & 2020 & 2021  \\ \hline
        Hagan SABR & 1.488$\times 10^{-2}$ & 1.180$\times 10^{-2}$ & 9.754$\times 10^{-3}$ & 8.922$\times 10^{-3}$  \\ \hline
        skew SABR & \textbf{3.279$\times \textbf{10}^{\textbf{-3}}$} & \textbf{3.017$\times \textbf{10}^{\textbf{-3}}$} & \textbf{2.472$\times \textbf{10}^{\textbf{-3}}$} & 2.730$\times 10^{-3}$  \\ \hline
        SVI & 5.929$\times 10^{-3}$ & 8.887$\times 10^{-3}$ & 7.988$\times 10^{-3}$ & 6.798$\times 10^{-3}$  \\ \hline
        cubic spline & 3.376$\times 10^{-3}$ & 3.135$\times 10^{-3}$ & 2.803$\times 10^{-3}$ & \textbf{2.566$\times \textbf{10}^{\textbf{-3}}$}  \\ \hline
        polynomial & 1.281$\times 10^{-2}$ & 4.511$\times 10^{-3}$ & 4.530$\times 10^{-3}$ & 4.461$\times 10^{-3}$  \\ \hline
        ~ & 2022 & 2023 & 2024 & all  \\ \hline
        Hagan SABR & 1.102$\times 10^{-2}$ & 5.050$\times 10^{-3}$ & 6.138$\times 10^{-3}$ & 9.340$\times 10^{-3}$  \\ \hline
        skew SABR & \textbf{2.527$\times \textbf{10}^{\textbf{-3}}$} & 1.669$\times 10^{-3}$ & \textbf{2.032$\times \textbf{10}^{\textbf{-3}}$} & \textbf{2.460$\times \textbf{10}^{\textbf{-3}}$}  \\ \hline
        SVI & 5.803$\times 10^{-3}$ & 4.348$\times 10^{-3}$ & 6.062$\times 10^{-3}$ & 6.484$\times 10^{-3}$  \\ \hline
        cubic spline & 2.675$\times 10^{-3}$ & \textbf{1.583$\times \textbf{10}^{\textbf{-3}}$} & 2.093$\times 10^{-3}$ & 2.537$\times 10^{-3}$  \\ \hline
        polynomial & 4.545$\times 10^{-3}$ & 1.613$\times 10^{-3}$ & 2.117$\times 10^{-3}$ & 4.746$\times 10^{-3}$  \\ \hline
    \end{tabular}
\end{table}

\begin{table}[H]
    \centering
 \caption{Model fitting results for ChiNext ETF (159915) options  (Oct 2022 -- Dec 2024)}
\renewcommand{\arraystretch}{1.5}
\tabcolsep=0.2cm
    \begin{tabular}{|c|c|c|c|c|}
\hline \makecell{159915 \\ (2022.10-2024.12)} & 2022 & 2023 & 2024 & all  \\ \hline
        Hagan SABR & 4.062$\times 10^{-3}$ & 9.649$\times 10^{-3}$ & 1.634$\times 10^{-2}$ & 1.259$\times 10^{-2}$  \\ \hline
        skew SABR & 1.794$\times 10^{-3}$ & \textbf{3.059$\times \textbf{10}^{\textbf{-3}}$} & \textbf{4.536$\times \textbf{10}^{\textbf{-3}}$} & \textbf{3.644$\times \textbf{10}^{\textbf{-3}}$}  \\ \hline
        SVI & 3.813$\times 10^{-3}$ & 6.359$\times 10^{-3}$ & 1.411$\times 10^{-2}$ & 1.091$\times 10^{-2}$  \\ \hline
        cubic spline & 1.822$\times 10^{-3}$ & 3.065$\times 10^{-3}$ & 4.634$\times 10^{-3}$ & 3.896$\times 10^{-3}$  \\ \hline
        polynomial & \textbf{1.776$\times \textbf{10}^{\textbf{-3}}$} & 5.838$\times 10^{-3}$ & 1.654$\times 10^{-2}$ & 1.067$\times 10^{-2}$  \\ \hline
    \end{tabular}
\end{table}

From the tables, the skew-SABR model achieves the lowest fitting error in most cases, indicating the best overall fitting performance. Only in a few instances do the cubic spline or polynomial methods slightly outperform the skew-SABR model; however, the differences in fitting errors are typically marginal, and their overall performance remains broadly comparable.

During periods of elevated volatility or heightened market uncertainty (e.g., 2024), the fitting errors of traditional parametric models, such as Hagan-SABR and SVI, tend to increase, in some cases reaching the order of $10^{-2}$. This deterioration becomes more pronounced when the implied volatility curve exhibits complex structures, such as the presence of outliers or significant variations in curvature. In such scenarios, the parametric forms of these traditional models are often insufficient to fully capture the true structure of market-implied volatility.

Finally, to provide a more intuitive comparison of model fitting performance, the ChiNext ETF option (159915) is selected,  and the fitting results of different models are visualized at two representative time points. The figures display the market-implied volatility across strike prices, together with the fitted curves obtained from the Hagan-SABR, skew-SABR, SVI, and polynomial models. Meanwhile, the interpolation result from the cubic spline method is used as a benchmark curve and is represented by a blue dashed line, serving as a reference to illustrate the deviation of parametric models relative to a data-driven approach.

\begin{figure}[htbp]
  \subfloat[]{\includegraphics[width=0.47\textwidth]{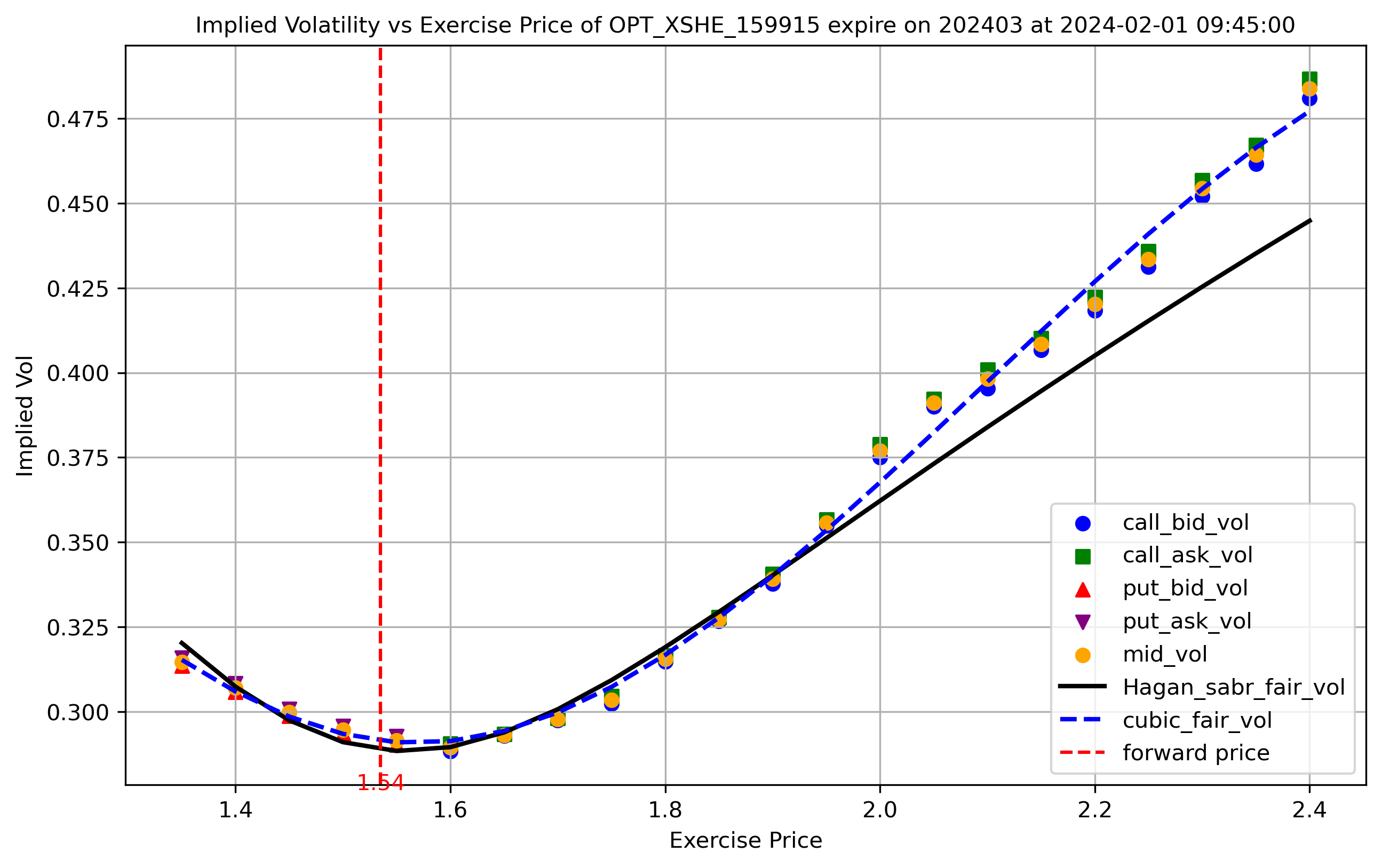}}
  \hfill
  \subfloat[]{\includegraphics[width=0.47\textwidth]{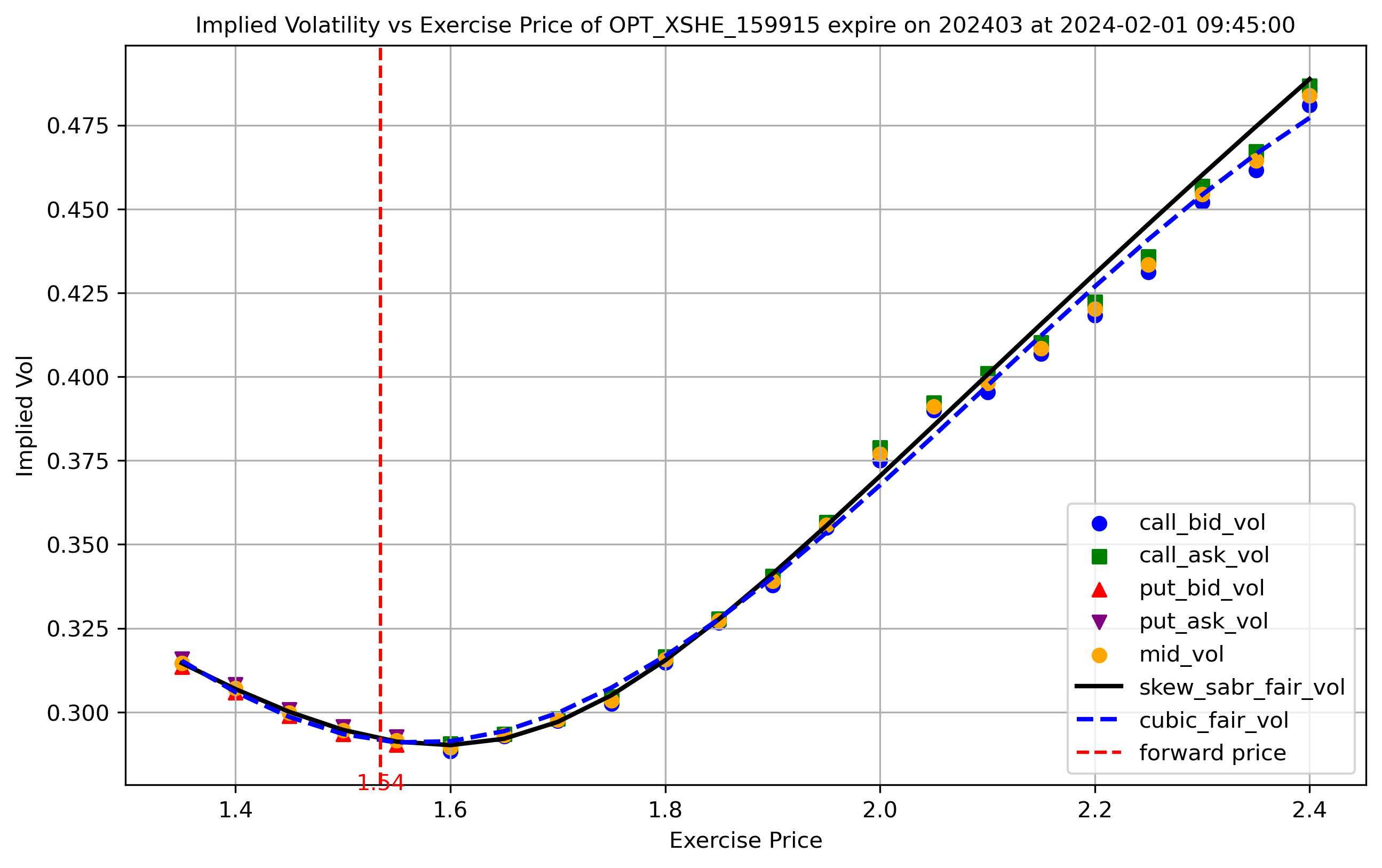}}
 
  \subfloat[]{\includegraphics[width=0.47\textwidth]{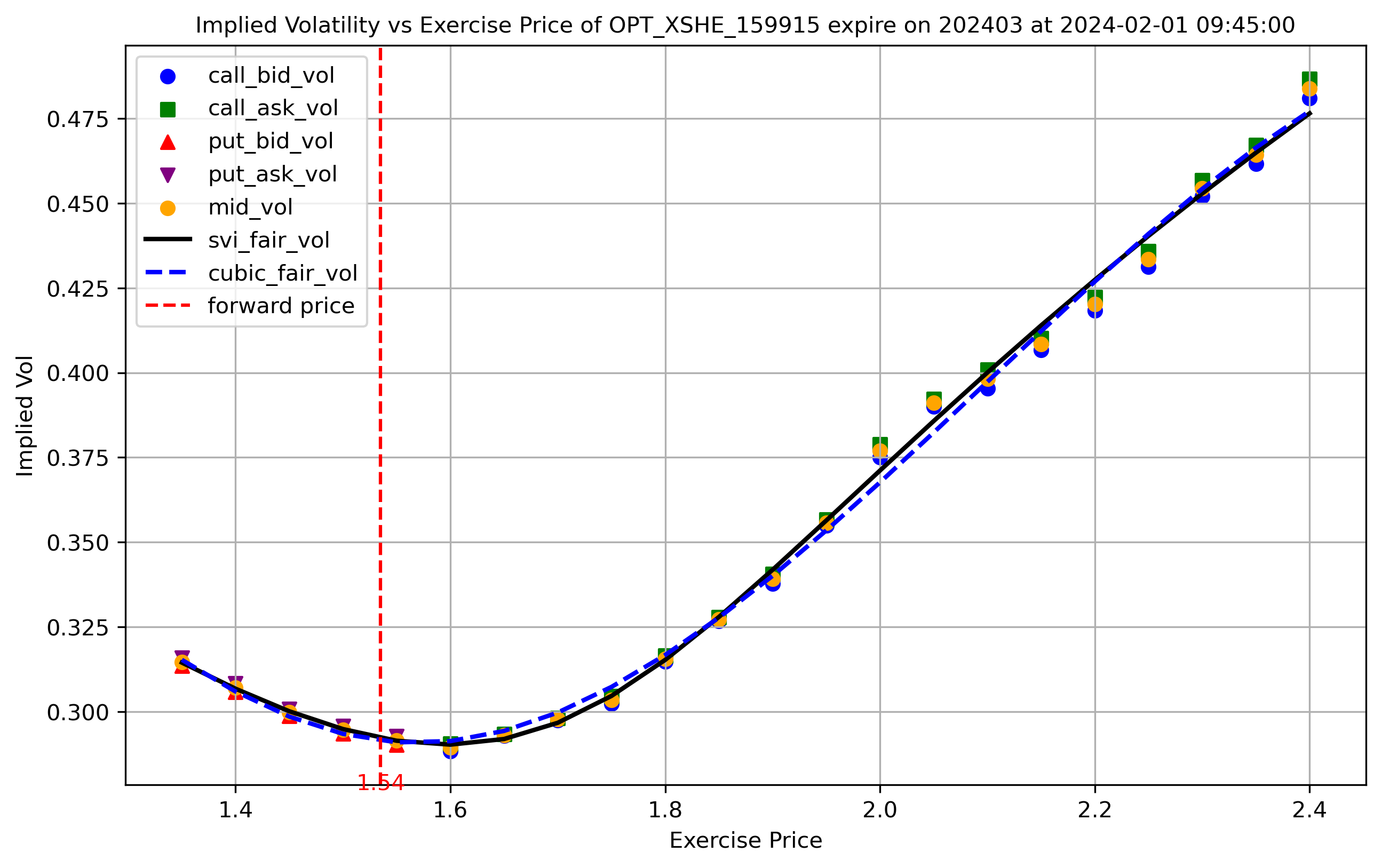}}
  \hfill
  \subfloat[]{\includegraphics[width=0.47\textwidth]{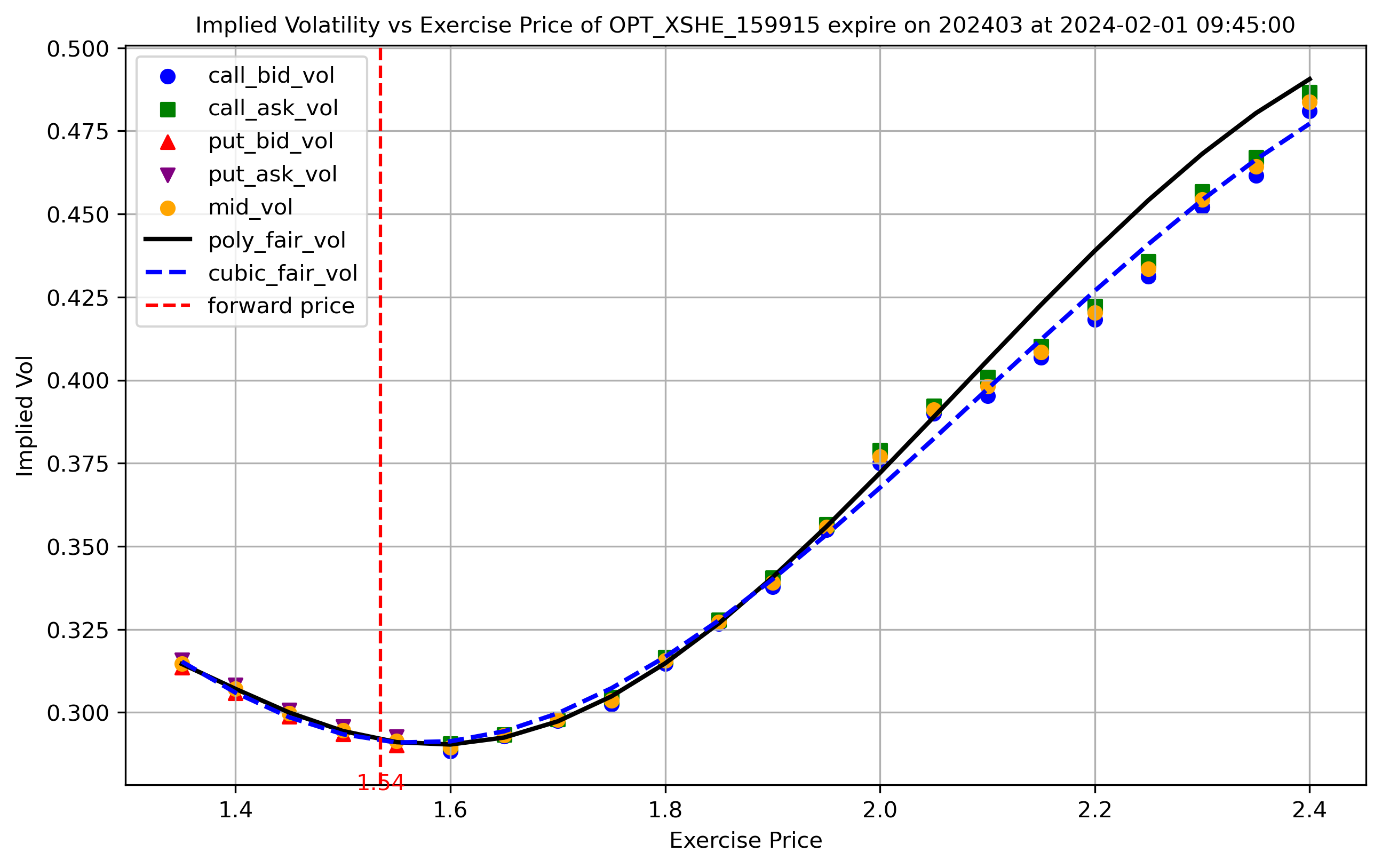}}
\caption{Fitting results of different models for ChiNext ETF option implied volatility curve (February 1, 2024, 9:45:00). (a). Hagan-SABR. (b). skew-SABR model. (c). SVI model. (d). Polynomial method.} \label{skew-fig28}
\end{figure}

First, consider the market at 9:45:00 on February 1, 2024. The implied volatility curve at this time exhibits an overall pattern of a downward slope on the left and an upward slope on the right, accompanied by several outliers, as shown in Fig.\ref{skew-fig28}. The fitting results indicate that both the skew-SABR and SVI models achieve relatively high accuracy for this curve and are able to capture the overall structure of the volatility curve effectively. In contrast, the Hagan-SABR model and the polynomial method exhibit relatively larger fitting errors. Specifically, the Hagan-SABR model performs poorly in the deep out-of-the-money region and fails to accurately capture the tail behavior of implied volatility; meanwhile, the polynomial model is sensitive to a few outliers on the right tail, leading to noticeable deviations in that region.

From a market perspective, the ChiNext ETF experienced a sustained decline over several trading days prior to this time point, causing the at-the-money strike to be relatively low compared to the full range of strikes. At the same time, the market formed expectations of a potential rebound, increasing the demand for out-of-the-money call options and thereby pushing up their implied volatility. This ultimately leads to the observed “low on the left, high on the right” volatility structure. Such asymmetric patterns are typically characterized in finance by the concept of skew. Since the skew-SABR model explicitly incorporates parameters that capture skewness, it is better suited to describe such asymmetric volatility distributions and therefore achieves superior fitting performance in such market environments.

\begin{figure}[htbp]
  \subfloat[]{\includegraphics[width=0.47\textwidth]{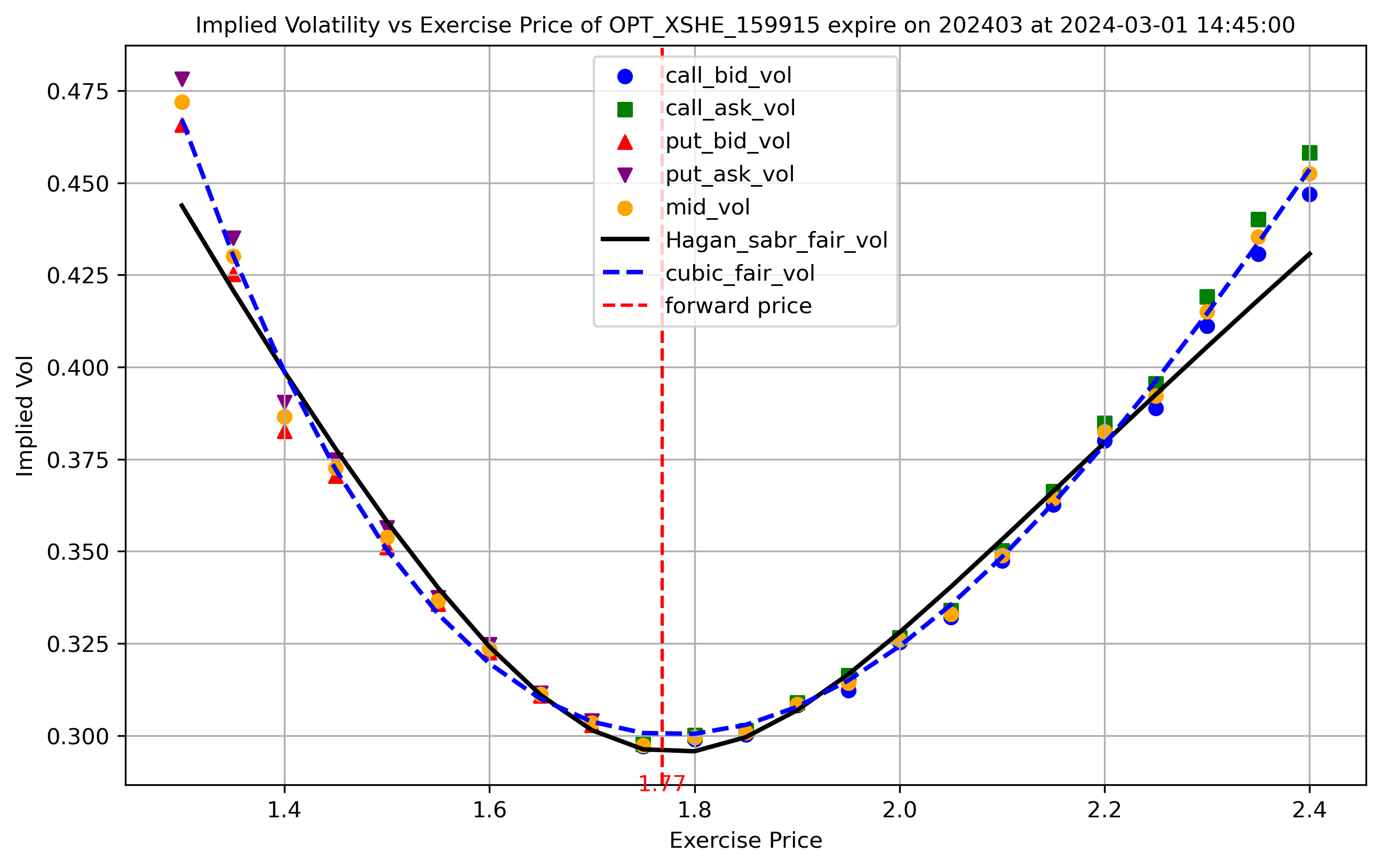}}
  \hfill
  \subfloat[]{\includegraphics[width=0.47\textwidth]{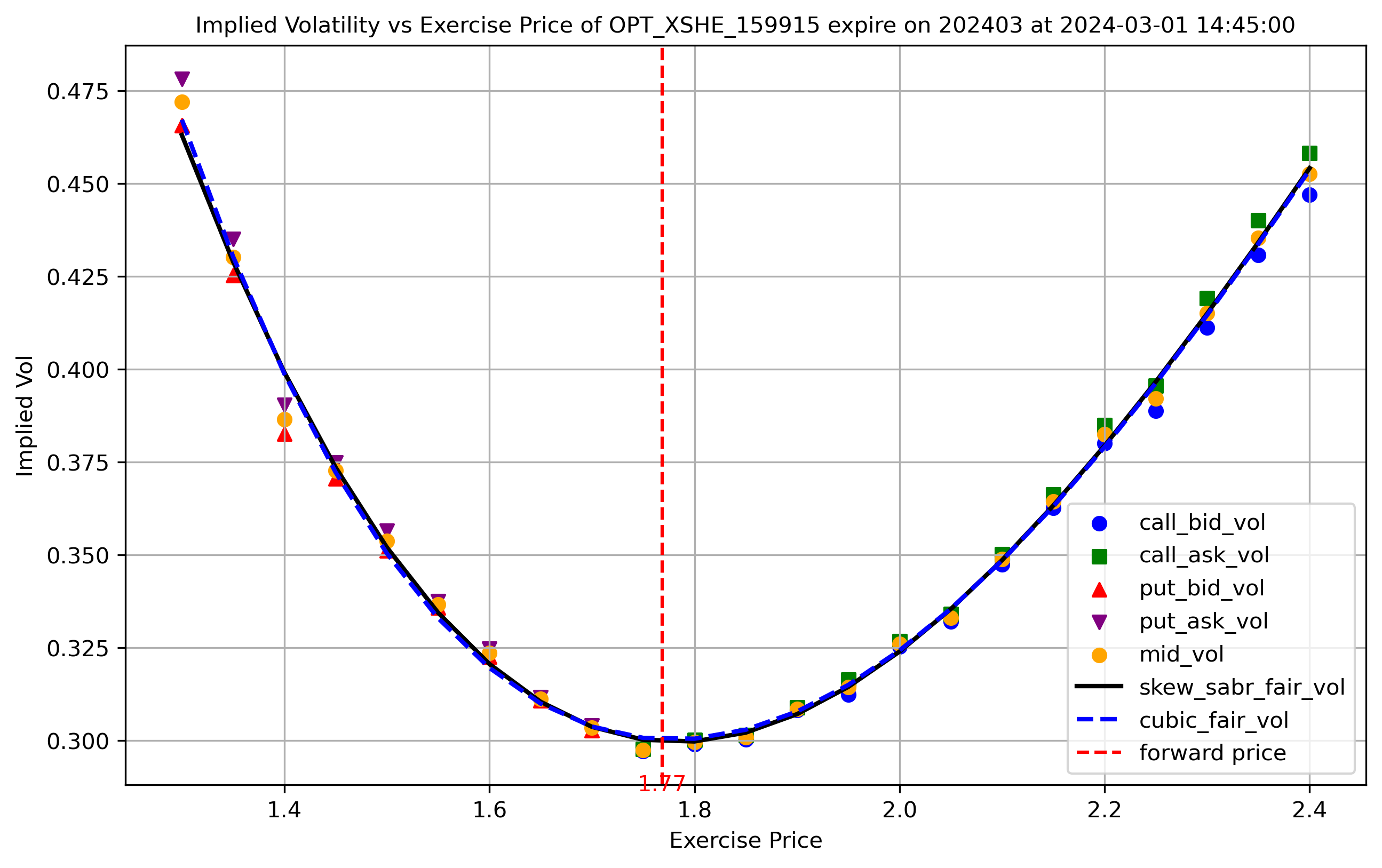}}
 
  \subfloat[]{\includegraphics[width=0.47\textwidth]{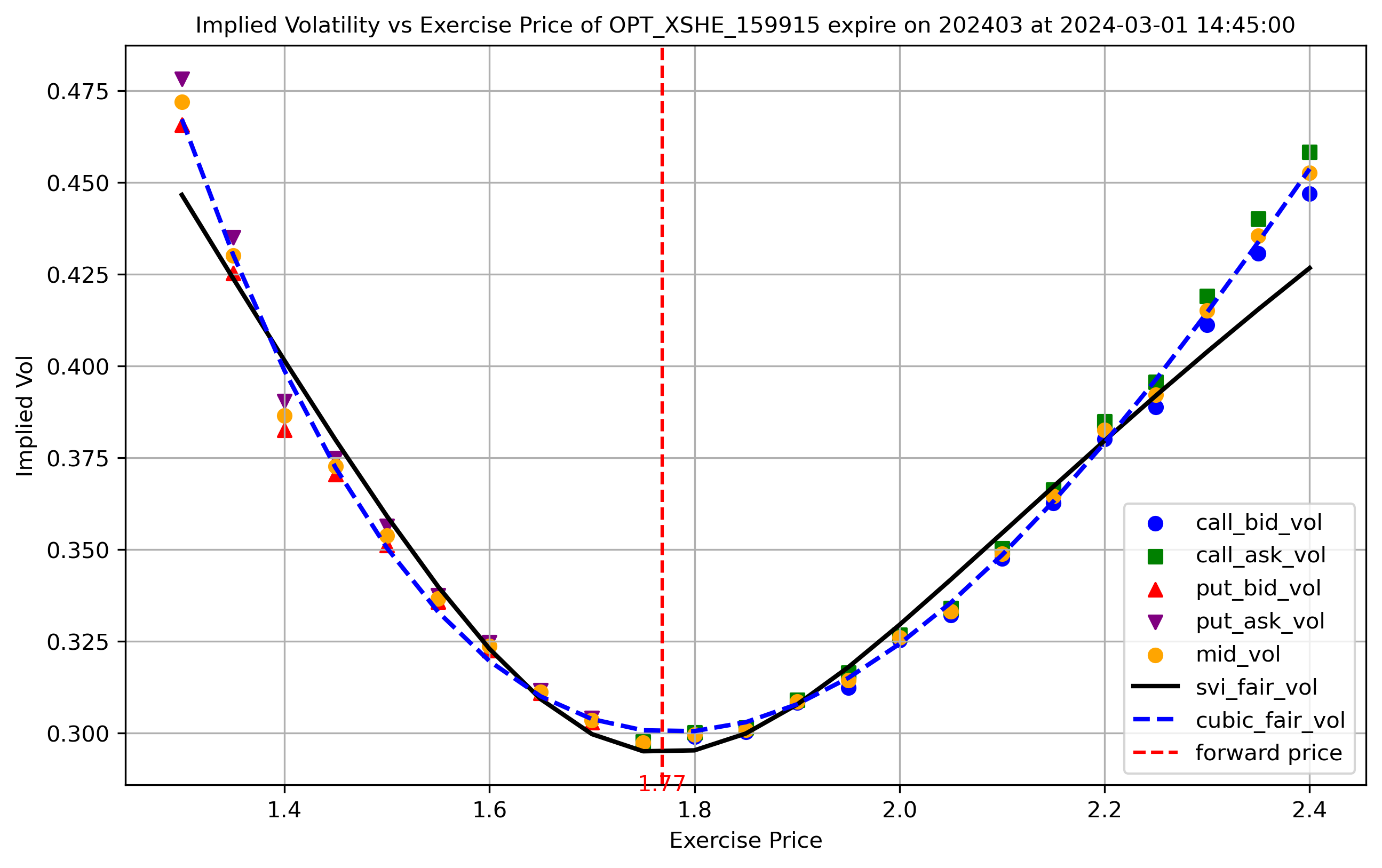}}
  \hfill
  \subfloat[]{\includegraphics[width=0.47\textwidth]{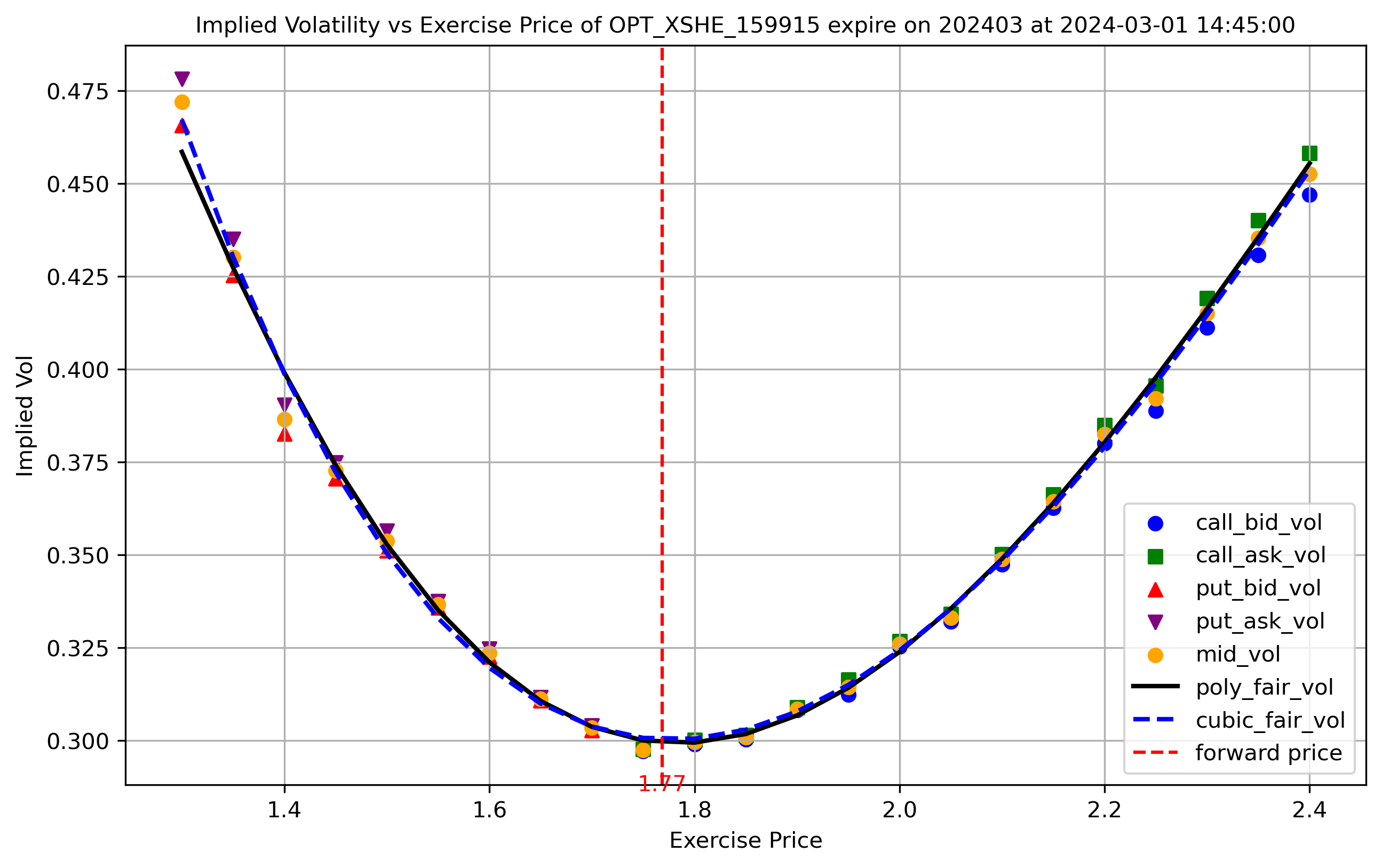}}
\caption{Fitting results of different models for ChiNext ETF option implied volatility curve (March 1, 2024, 14:45:00). (a). Hagan-SABR. (b). skew-SABR model. (c). SVI model. (d). Polynomial method.} \label{skew-fig32}
\end{figure}

The second time point is selected as March 1, 2024, at 14:45:00. At this moment, the implied volatility curve is approximately centered around the at-the-money (ATM) region, with the overall volatility levels on both sides being relatively similar, as shown in Fig.~\ref{skew-fig32}. However, in contrast to the previous timestamp, the curvature on the two sides of the curve exhibits a noticeable difference, i.e., the degree of convexity differs between the left and right wings, resulting in a pronounced higher-order asymmetric structure. This feature is not merely a first-order skewness effect, but is instead reflected in the asymmetric rates of change in curvature across different directions of log-moneyness.

Under such circumstances, the fitting capability of the SVI model is subject to certain limitations. Since SVI controls the overall curvature through a single parameter (namely $\sigma$ in Eq.~(\ref{skew-11})), it is difficult for the model to simultaneously match the differing curvature dynamics on both sides. As a result, the model tends to exhibit relatively larger fitting errors in curves with pronounced higher-order asymmetry. In contrast, polynomial-based approaches, by explicitly incorporating higher-order terms, offer greater functional flexibility in capturing such curvature differences, thereby achieving better fitting performance for this type of cross-section. However, polynomial models lack structural constraints, and their parameters generally do not carry clear financial interpretations.

By comparison, the skew-SABR model demonstrates stronger robustness in such market conditions. Its advantage lies in modeling the volatility of variance, which allows the implied volatility in the wings to adjust nonlinearly with respect to log-strike. Meanwhile, the interaction between skew-related parameters and volatility dynamics provides a structural mechanism for generating asymmetric curvature across the two sides. Therefore, when the implied volatility curve exhibits strong higher-order asymmetry, the skew-SABR model is capable of naturally capturing the asymmetric wing behavior within a unified functional form, without the need to introduce piecewise functions or additional higher-order polynomial terms.

Synthesizing the analysis of the two representative time points, the curve at the former timestamp primarily exhibits a first-order skew asymmetric structure, while the latter 
presents more complex high-order asymmetric characteristics.
Under these two distinct types of asymmetry, the Hagan-SABR, SVI, and polynomial models all suffer from certain structural limitations, which lead to varying degrees of fitting bias under specific market conditions. In contrast, the skew-SABR model is able to  adapt more consistently to changes in the cross-sectional shape of the volatility surface, demonstrating greater flexibility and robustness across different curve configurations.

Further, by combining the tabulated results and visual analysis presented above, it can be observed that across different market regimes—including periods of significantly elevated volatility and relatively stable phases—as well as under various asymmetric volatility structures, the skew-SABR model consistently achieves high and stable fitting accuracy. In most cases, its error level is comparable to that of the cubic spline method, and in some instances even lower. Notably, in short-term maturities close to expiration, where other parametric models tend to exhibit significantly increased fitting errors, the skew-SABR model remains relatively stable in its performance.

In addition, compared with the purely numerical interpolation-based cubic spline method, the parameters of the skew-SABR model possess clear financial interpretations. This enables the model to maintain strong interpretability while achieving high fitting accuracy. Overall, the above numerical results suggest that the skew-SABR model not only effectively captures skewness and higher-order asymmetric structures in implied volatility curves, but also delivers stable and interpretable fitting performance across different market environments and maturity structures.

\section{Conclusion and Future Work}

\label{skew-sec5}

This paper extends the stochastic processes of the underlying asset price and its volatility within the classic Hagan-SABR framework, and proposes the skew-SABR model to more flexibly capture asymmetric structures in implied volatility curves and improve fitting accuracy. 
From a theoretical perspective, under the newly specified stochastic dynamics, an explicit expression for the Black implied volatility is derived and further simplified into a form that closely resembles the Hagan-SABR formula while explicitly incorporating skewness. In addition, the consistency of the model specification and the correctness of the derivation are systematically validated, and the financial interpretations of the model parameters are clearly established. 
From an empirical perspective, using data from the Chinese options market spanning 2018 to 2025, the skew-SABR model is systematically compared with the Hagan-SABR, SVI, polynomial, and spline-based methods. The results show that the skew-SABR model achieves the best or near-best fitting accuracy in the vast majority of market conditions, and demonstrates strong robustness, particularly when the implied volatility surface exhibits pronounced asymmetry. Overall, the skew-SABR model preserves the parsimony of the SABR framework while substantially enhancing its ability to capture complex volatility structures, thereby combining theoretical rigor with practical applicability.

Although the skew-SABR model proposed in this paper achieves satisfactory performance in fitting implied volatility curves, there remains scope for further research and extension.  
Future work may proceed along the following directions. First, the model can be extended to multi-maturity settings to analyze the dynamic evolution of parameters across the time dimension. Second, the performance of the skew-SABR model can be further investigated in practical financial applications such as option pricing, risk management, and volatility forecasting, in order to provide a more comprehensive assessment of its practical effectiveness and robustness.

\section{Acknowledgments}

This research was funded by 
the Strategic Priority Research Program of Chinese Academy of
Sciences (Grant No. XDB0500000) and the National Natural Science Foundation of China (Grant
No. 12371413). 
The data used in this paper are sourced from  the International Digital Economy Academy.

\bibliographystyle{unsrt}
\bibliography{references}

\end{document}